\documentclass[ALICE,manyauthors]{cernphprep}
\usepackage[comma,square,numbers,sort&compress]{natbib}

\usepackage{lineno}
\usepackage{xspace}
\usepackage{hyperref}
\usepackage{xcolor}
\usepackage[section]{placeins} 
\usepackage[T1]{fontenc}
\usepackage{orcidlink}

\begin{document}
%

\newcommand{\pp}           {pp\xspace}
\newcommand{\ppbar}        {\mbox{$\mathrm {p\overline{p}}$}\xspace}
\newcommand{\XeXe}         {\mbox{Xe--Xe}\xspace}
\newcommand{\PbPb}         {\mbox{Pb--Pb}\xspace}
\newcommand{\pA}           {\mbox{p--A}\xspace}
\newcommand{\pPb}          {\mbox{p--Pb}\xspace}
\newcommand{\AuAu}         {\mbox{Au--Au}\xspace}
\newcommand{\dAu}          {\mbox{d--Au}\xspace}

\newcommand{\s}            {\ensuremath{\sqrt{s}}\xspace}
\newcommand{\snn}          {\ensuremath{\sqrt{s_{\mathrm{NN}}}}\xspace}
\newcommand{\pt}           {\ensuremath{p_{\rm T}}\xspace}
\newcommand{\meanpt}       {\ensuremath{\langle p_{\mathrm{T}}\rangle}\xspace}
\newcommand{\meanptsq}       {\ensuremath{\langle p_{\mathrm{T}}^{\mathrm{2}}\rangle}\xspace}
\newcommand{\ycms}         {\ensuremath{y_{\rm CMS}}\xspace}
\newcommand{\ylab}         {\ensuremath{y_{\rm lab}}\xspace}
\newcommand{\y}         {\ensuremath{y}\xspace}
\newcommand{\etarange}[1]  {\mbox{$\left | \eta \right |~<~#1$}}
\newcommand{\yrange}[1]    {\mbox{$\left | y \right |<#1$}}

\newcommand{\rapidity} {\ensuremath{y}\xspace}
\newcommand{\dndy}         {\ensuremath{\mathrm{d}N/\mathrm{d}y}\xspace}
\newcommand{\dndeta}       {\ensuremath{\mathrm{d}N_\mathrm{ch}/\mathrm{d}\eta}\xspace}
\newcommand{\avdndeta}     {\ensuremath{\langle\dndeta\rangle}\xspace}
\newcommand{\dNdy}         {\ensuremath{\mathrm{d}N_\mathrm{ch}/\mathrm{d}y}\xspace}
\newcommand{\dndydpt}         {\ensuremath{\mathrm{d^2}N/(\mathrm{d}y\mathrm{d}p_{\mathrm{T}}}\xspace)}
\newcommand{\Npart}        {\ensuremath{N_\mathrm{part}}\xspace}
\newcommand{\TAA}          {\ensuremath{T_{\text{AA}}}}
\newcommand{\Ncoll}        {\ensuremath{N_\mathrm{coll}}\xspace}
\newcommand{\dEdx}         {\ensuremath{\textrm{d}E/\textrm{d}x}\xspace}
\newcommand{\RpPb}         {\ensuremath{R_{\rm pPb}}\xspace}
\newcommand{\RAA}         {\ensuremath{R_{\rm AA}}\xspace}
\newcommand{\rAA}         {\ensuremath{r_{\rm AA}}\xspace}
\newcommand{\AccEff}      {\ensuremath{{\rm A}\times\epsilon}\xspace}
\newcommand{\mee}         {\ensuremath{m_{\rm e^+e^-}}\xspace}
\newcommand{\ccbar} 
{\ensuremath{\mathrm{c}\rm\overline{c}}\xspace}

\newcommand{\nineH}        {$\sqrt{s}~=~0.9$~Te\kern-.1emV\xspace}
\newcommand{\seven}        {$\sqrt{s}~=~7$~Te\kern-.1emV\xspace}
\newcommand{\twoH}         {$\sqrt{s}~=~0.2$~Te\kern-.1emV\xspace}
\newcommand{\twosevensix}  {$\sqrt{s}~=~2.76$~Te\kern-.1emV\xspace}
\newcommand{\five}         {$\sqrt{s}~=~5.02$~Te\kern-.1emV\xspace}
\newcommand{\twosevensixnn}{$\sqrt{s_{\mathrm{NN}}}~=~2.76$~Te\kern-.1emV\xspace}
\newcommand{\fivenn}       {$\sqrt{s_{\mathrm{NN}}}~=~5.02$~Te\kern-.1emV\xspace}
\newcommand{\LT}           {L{\'e}vy-Tsallis\xspace}
\newcommand{\GeVc}         {Ge\kern-.1emV/$c$\xspace}
\newcommand{\MeVc}         {Me\kern-.1emV/$c$\xspace}
\newcommand{\TeV}          {Te\kern-.1emV\xspace}
\newcommand{\GeV}          {Ge\kern-.1emV\xspace}
\newcommand{\MeV}          {Me\kern-.1emV\xspace}
\newcommand{\GeVmass}      {Ge\kern-.2emV/$c^2$\xspace}
\newcommand{\MeVmass}      {Me\kern-.2emV/$c^2$\xspace}
\newcommand{\lumi}         {\ensuremath{\mathcal{L}}\xspace}

\newcommand{\ITS}          {\rm{ITS}\xspace}
\newcommand{\TOF}          {\rm{TOF}\xspace}
\newcommand{\ZDC}          {\rm{ZDC}\xspace}
\newcommand{\ZDCs}         {\rm{ZDCs}\xspace}
\newcommand{\ZNA}          {\rm{ZNA}\xspace}
\newcommand{\ZNC}          {\rm{ZNC}\xspace}
\newcommand{\SPD}          {\rm{SPD}\xspace}
\newcommand{\SDD}          {\rm{SDD}\xspace}
\newcommand{\SSD}          {\rm{SSD}\xspace}
\newcommand{\TPC}          {\rm{TPC}\xspace}
\newcommand{\TRD}          {\rm{TRD}\xspace}
\newcommand{\VZERO}        {\rm{V0}\xspace}
\newcommand{\VZEROA}       {\rm{V0A}\xspace}
\newcommand{\VZEROC}       {\rm{V0C}\xspace}
\newcommand{\Vdecay} 	   {\ensuremath{V^{0}}\xspace}

\newcommand{\ee}           {\ensuremath{\rm{e}^{+}\rm{e}^{-}}\xspace}
\newcommand{\mumu}           {\ensuremath{\rm{\mu}^{+}\rm{\mu}^{-}}\xspace} 
\newcommand{\eerange}           {\ensuremath{ ~<~ e^{+}e^{-} ~<~}}

\newcommand{\pip}          {\ensuremath{\pi^{+}}\xspace}
\newcommand{\pim}          {\ensuremath{\pi^{-}}\xspace}
\newcommand{\kap}          {\ensuremath{\rm{K}^{+}}\xspace}
\newcommand{\kam}          {\ensuremath{\rm{K}^{-}}\xspace}
\newcommand{\pbar}         {\ensuremath{\rm\overline{p}}\xspace}
\newcommand{\kzero}        {\ensuremath{{\rm K}^{0}_{\rm{S}}}\xspace}
\newcommand{\dzero}        {\ensuremath{{\rm D}^{0}}\xspace}
\newcommand{\lmb}          {\ensuremath{\Lambda}\xspace}
\newcommand{\almb}         {\ensuremath{\overline{\Lambda}}\xspace}
\newcommand{\Om}           {\ensuremath{\Omega^-}\xspace}
\newcommand{\Mo}           {\ensuremath{\overline{\Omega}^+}\xspace}
\newcommand{\X}            {\ensuremath{\Xi^-}\xspace}
\newcommand{\Ix}           {\ensuremath{\overline{\Xi}^+}\xspace}
\newcommand{\Xis}          {\ensuremath{\Xi^{\pm}}\xspace}
\newcommand{\Oms}          {\ensuremath{\Omega^{\pm}}\xspace}
\newcommand{\degree}       {\ensuremath{^{\rm o}}\xspace}
\newcommand{\jpsi}         {\ensuremath{{\rm J}/\psi}\xspace}
\newcommand{\chndf}       {\ensuremath{\chi^{2}/N_{dof}}\xspace}      
\newcommand{\psitwos}         {\ensuremath{\psi(\rm 2S)}\xspace}
\newcommand{\todo}[1]{{\textcolor{red}{\bf [#1]}}}

\begin{titlepage}
\PHyear{2023}       
\PHnumber{054}      
\PHdate{22 March}  

\title{Measurements of inclusive \textbf{J/$\psi$} production at midrapidity and forward rapidity in \PbPb collisions at \textbf{$\sqrt{s_{\mathrm{NN}}}$ = 5.02 TeV}}
\ShortTitle{Inclusive J$/\psi$ production in Pb--Pb collisions at \fivenn}   

\Collaboration{ALICE Collaboration\thanks{See Appendix~\ref{app:collab} for the list of collaboration members}}
\ShortAuthor{ALICE Collaboration} 

\begin{abstract}
The measurements of the inclusive \jpsi yield at midrapidity (\yrange0.9) and forward rapidity (2.5 $< y <$ 4) in \PbPb collisions at \fivenn with the ALICE detector at the LHC are reported.
The inclusive \jpsi production yields and nuclear modification factors, \RAA, are measured as a function of the collision centrality, \jpsi transverse momentum (\pt), and rapidity. The \jpsi average transverse momentum and squared transverse momentum (\meanpt and \meanptsq) are evaluated as a function of the centrality at midrapidity. Compared to the previous ALICE publications, here the entire \PbPb collisions dataset collected during the LHC Run 2 is used, which improves the precision of the measurements and extends the \pt coverage. The \pt-integrated \RAA shows a hint of an increasing trend towards unity from semicentral to central collisions at midrapidity, while it is flat at forward rapidity. The \pt-differential \RAA shows a strong suppression at high \pt with less suppression at low \pt where it reaches a larger value at midrapidity compared to forward rapidity. The ratio of the \pt-integrated yields of \jpsi to those of D$^{0}$ mesons is reported for the first time for the central and semicentral event classes at midrapidity. Model calculations implementing charmonium production via the coalescence of charm quarks and antiquarks during the fireball evolution (transport models) or in a statistical approach with thermal weights are in good agreement with the data at low \pt. At higher \pt, the data are well described by transport models and a model based on energy loss in the strongly-interacting medium produced in nuclear collisions at the LHC.

\end{abstract}
\end{titlepage}

\setcounter{page}{2} 


\section{Introduction} \label{sec:Introduction}
Quantum chromodynamics (QCD) is the theory describing the strong interaction. Lattice QCD, i.e. the discrete formulation of QCD, predicts the existence of a state of deconfined matter at high energy density that is characterised by quark and gluon degrees of freedom~\cite{Bazavov:2009zn,Borsanyi:2013bia}. This quark--gluon plasma (QGP) is created during the early hot and dense stage of heavy-ion collisions at ultra-relativistic energies. 

The heavy quarks, charm (c) and beauty (b), are unique probes for this phase of matter~\cite{Prino:2016cni,Rothkopf:2019ipj}. Due to their large masses, they are produced as quark--antiquark pairs in hard partonic scattering processes in the early stage of the collision, and they thus experience the full evolution of the system. While the majority of the produced heavy quarks and antiquarks hadronise independently into open heavy-flavour hadrons, bound quarkonium states can also be formed~\cite{Brambilla:2010cs}. However, it has been predicted that the formation of bound states should be suppressed due to the mechanism of colour screening where the large density of colour charges in the QGP hinders the production of bound quarkonia~\cite{Matsui:1986dk,Digal:2001ue}. The degree of suppression of the various quarkonium states depends on their binding energy along with the medium properties, such as its temperature. Consequently, the measurements of quarkonium production rates in heavy-ion collisions have been considered as a potential thermometer of the medium.

The production of \jpsi, the charmonium ground state with quantum numbers $J^{PC} = 1^{--}$, has been studied extensively in heavy-ion collisions over the last several decades. Suppression of the \jpsi yield was observed in nucleus--nucleus collisions with respect to the expectation from proton--proton collisions at the SPS (up to centre-of-mass energy per nucleon pair \snn $\mathrm{=}$ 17~GeV)~\cite{NA38:1994yau,NA50:2004sgj,NA60:2007ewx}, at RHIC (\snn up to 200~GeV)~\cite{PHENIX:2006gsi,PHENIX:2011img,STAR:2009irl,STAR:2013eve} and at the LHC (\snn $\mathrm{=}$ 2.76~TeV~\cite{CMS:2012bms,ALICE:2012jsl,ALICE:2013osk,ALICE:2015jrl} and 5.02~TeV~\cite{ALICE:2016flj,ALICE:2019lga, ALICE:2019nrq,ALICE:2022wpn, ATLAS:2018hqe,CMS:2017uuv}). However, contrary to the prediction from the colour-screening scenario, the measured suppression does not increase with increasing collision energy from RHIC to LHC despite of an increased energy density of the produced QGP. At LHC energies, \jpsi production is found to be less suppressed than at the lower RHIC energies, in particular at low transverse momentum, \pt, of the \jpsi~\cite{ALICE:2015jrl,starjpsi39-200,PhysRevLett.101.122301,PhysRevC.93.034903}. In addition, a significant azimuthal anisotropy in the \jpsi production was observed via the elliptic and triangular flow measurements reported in Refs.~\cite{ALICE:2017quq,ALICE:2020pvw}. These observations are explained by an additional \jpsi production mechanism, referred to as (re)generation in the following, in which copiously produced uncorrelated charm quarks and antiquarks bind into \jpsi mesons~\cite{Braun-Munzinger:2000csl,Thews:2000rj}. This process can only take place in a deconfined medium, and its contribution to the measured \jpsi yield increases with the density of c$\overline{\rm c}$ pairs and, therefore, with increasing collision energy and decreasing \pt~\cite{Andronic:2019wva,Zhou:2014kka,Du:2015wha}. With increasing \pt, (re)generation becomes less relevant for \jpsi production and, instead, charmonium dissociation and the fragmentation of high-energy partons into charmonia become dominant. In the latter case, the suppression of high-\pt \jpsi yields should reflect the energy loss of partons~\cite{Arleo:2017ntr}, which is mostly of radiative nature in this kinematic regime.

The formation process of charmonia in heavy-ion collisions is complex and various phenomenological approaches are considered. In the statistical hadronization scenario, the relative abundances of charmomium states with respect to other charmed hadrons  are determined by thermal weights~\cite{Braun-Munzinger:2000csl,Andronic:2019wva} at the system chemical freeze-out. In microscopic transport and comover interaction models, charmonia are continuously produced and broken up during their propagation through the QGP~\cite{Thews:2000rj,Zhou:2014kka,Du:2015wha,Ferreiro:2012rq}. 
Furthermore, it is important to consider cold nuclear matter (CNM) effects. In particular, the modification of the parton distribution functions in nuclei with respect to nucleons~\cite{Armesto:2006ph} has to be taken into account for the interpretation of the results. 
These CNM effects were investigated in ALICE especially with proton--nucleus collisions~\cite{ALICE:2013snh,ALICE:2015sru, ALICE:2015kgk,ALICE:2018mml,ALICE:2018szk, ALICE:2020tsj, ALICE:2021lmn}.

For a better assessment of the production mechanisms, systematic measurements of the centrality, \pt, and rapidity dependence of \jpsi production are pivotal. In this article, the ALICE results on inclusive \jpsi production at midrapidity (\yrange0.9) for $0.15 < \pt < 15$~GeV/c and forward rapidity (2.5 $< y <$ 4) for $0.3 < \pt < 20$~GeV/c, from the full Run~2 data sample at the LHC, are reported. Inclusive \jpsi measurements contain a prompt \jpsi contribution from direct \jpsi and decay from heavier charmonium states, and a non-prompt \jpsi contribution from the decay of beauty hadrons. The precision of the measurements, using the entire Run 2 data sample, improved significantly compared to previous ones~\cite{ALICE:2019nrq,ALICE:2019lga} and the measurements could be extended up to 15~GeV/c and 20~GeV/c at mid and forward rapidity, respectively. The \pt-differential \jpsi yields in \PbPb collisions at \fivenn are measured in various centrality classes. At midrapidity, the average transverse momentum $\langle\pt\rangle$ as well as squared transverse momentum $\langle\pt^{2}\rangle$ is determined, which provide a quantitative estimation of the evolution of the \pt spectra as a function of centrality. The nuclear modification factor \RAA defined, as the ratio of the yield in \PbPb to the corresponding yield in pp collisions scaled by the number of binary nucleon--nucleon collisions is calculated. The results as a function of \pt and collision centrality are compared with model calculations employing the statistical hadronisation~\cite{Andronic:2019wva}, microscopic parton transport~\cite{Zhou:2014kka,Du:2015wha}, comover~\cite{Ferreiro:2012rq}, and energy loss~\cite{Arleo:2017ntr} approaches.

\section{Apparatus and data sample}
\label{detectors}  

A complete description of the ALICE apparatus and its performance is given in Refs.~\cite{Aamodt:2008zz, Abelev:2014ffa}. The central barrel detectors, for the dielectron analysis, and the muon spectrometer, for the dimuon analysis, covering midrapidity and forward rapidity, respectively, were used in the analyses reported in this paper. 

At midrapidity, the main detectors employed in the analysis are the Time Projection Chamber (TPC)~\cite{TPC} and the Inner Tracking System (ITS)~\cite{ITS}, both immersed in a uniform magnetic field of 0.5~T provided by a solenoid magnet. 
The TPC is used for tracking and particle identification. It covers the pseudorapidity range \etarange0.9 for tracks with full radial length and has full coverage in azimuth. It provides excellent momentum resolution and electron--hadron separation in a wide range of track transverse momentum.
The ITS is a cylindrical six-layer silicon detector, with the innermost layer located at 3.9 cm from the beam pipe, providing additional space points for tracking that enhance the spatial resolution in the reconstruction of primary and secondary vertices. 

 At forward rapidity, the muon spectometer~\cite{CERN-LHCC-99-022, CERN-LHCC-2000-046} detects muons in the range $-4 < \eta < -2.5$. It consists of a 3 Tm dipole associated with a tracking and a trigger system. A front absorber with a thickness of 10 interaction lengths is placed before the tracking system in order to filter out hadrons produced in the interaction. The tracking system consists of five tracking stations, each one made of two planes of cathode pad chambers. An iron wall with 7.2 interaction length thickness is located between the tracking and trigger stations in order to stop secondary hadrons escaping the front absorber and low momentum muons produced predominantly from $\pi$ and $K$ decays. The trigger system consists of two stations, each one made of two planes of resistive plate chambers. Finally, a conical absorber around the beam pipe protects the spectrometer against secondary particles produced by the interaction of primary particles with large pseudorapidity at the beam pipe. In the analysis at forward rapidity, the determination of the primary vertex of the collision is provided by the Silicon Pixel Detector (SPD) that constitutes the two innermost layers of the ITS.

Both analyses, at midrapidity and forward rapidity, use the V0~\cite{ALICE:2013axi} and Zero Degree Calorimeter (ZDC)~\cite{ALICE:2012aa} detectors. The V0 detector consists of two scintillator detector arrays and covers the full azimuth in the pseudorapidity regions $-3.7 < \eta < -1.7$ and 2.8 $< \eta < $ 5.1, respectively. It is used for triggering, beam--gas background rejection, and characterisation of the event centrality. The ZDC detectors are located at a distance of 112.5 m on both sides of the interaction region along the beam direction, and they detect spectator nucleons emitted at zero degree with respect to the LHC beam axis. They are used to reject electromagnetic \PbPb interactions. 

The trigger for minimum-bias (MB) events was provided by the coincidence of signals in the two scintillator arrays of the V0 detector. The dimuon analysis relies on a dimuon trigger which requires, in addition to the MB trigger, the detection of two opposite-sign tracks in the muon trigger system. The muon trigger selects muon candidates with a \pt larger than a threshold of $\sim 1$~\GeVc. The trigger efficiency reaches 50\% at this threshold value and a plateau value of 98\% at $\pt \sim 2.5$~\GeVc~\cite{Bossu:2012jt}.

The results presented in this article are based on the data sample collected by ALICE from Pb--Pb collisions at \fivenn in 2015 and 2018 during Run 2 at the LHC. During the 2018 data taking, the \PbPb dataset for the central barrel was enhanced with central (0--10\%) and semicentral (30--50\%) events. In total, the integrated luminosity corresponding to the analysed data sample was about~105 $\rm\mu \rm b^{-1}$ and~51 $\rm\mu \rm b^{-1}$ for the central and semicentral events, respectively. For the other centrality intervals, the integrated luminosity of the data sample was~22 $\rm\mu \rm b^{-1}$. For the analysis at forward rapidity, the dimuon triggered sample corresponds to an integrated luminosity of~756~$\rm\mu b^{-1}$. 
At midrapidity, triggered events containing collisions that overlap within a time window smaller than the readout time of the TPC were removed to preserve a uniform particle identification performance of the TPC, which is sensitive to the total charge produced by the ionising tracks in the sensitive volume. Only events with the primary vertex, reconstructed within $\pm$10~cm from the nominal interaction point in the beam direction, were considered for further analysis at midrapidity. In the forward analysis, there was no selection on the primary vertex. 

\section{Analysis details}
\label{sec:analysis} 

The primary observable  is the $\pt$-differential \jpsi yield per unit of rapidity $\mathrm{d}^2N/(\mathrm{d}y \; \mathrm{d}\pt)$. For a given interval of centrality, rapidity ($\mathrm{\Delta}y$), and transverse momentum ($\mathrm{\Delta}\pt$), this is obtained as
\begin{equation}
\frac{\mathrm{d}^{2} N}{\mathrm{~d} y \mathrm{~d} \pt} = \frac{N_{\jpsi}}{N_{\rm{ev}} \times \mathrm{BR}_{\jpsi \rightarrow \mathrm{l}^+\mathrm{l}^-} \times (A \times \epsilon) \times \Delta y \times \Delta \pt},
\label{eq:d2ndydpt}
\end{equation}
where $N_{\jpsi}$ is the number of reconstructed \jpsi mesons, $N_{\mathrm{ev}}$ is the number of events corresponding to the analysed centrality interval, and ($A\times\epsilon$) is the acceptance times efficiency factor. The branching ratio (BR) corresponds to either the dielectron or the dimuon \jpsi decay channel. 
Since the analysis at forward rapidity was based on a sample of dimuon-triggered events, the equivalent $N_{\mathrm{ev}}$ was obtained as the product of the number of dimuon-triggered events times the inverse of the probability of having a dimuon trigger in a MB triggered event, $F_{\mathrm{norm}}$~\cite{ALICE:2015kgk,ALICE:2020jff}. The number of equivalent $N_{\mathrm{ev}}$ was first obtained for the 0--90\% centrality class and was then scaled to the centrality classes considered in the analysis. 

The nuclear modification factor, \RAA, is obtained as
\begin{equation}
\RAA=\frac{\mathrm{d}^{2} N / (\mathrm{d} y \mathrm{~d} \pt)}{\langle T_{\mathrm{AA}} \rangle  \, \mathrm{d}^{2} \sigma_{\mathrm{pp}} / (\mathrm{d} y \mathrm{~d} \pt)},
\label{eq:Raa}
\end{equation}
where $\langle$\TAA$\rangle$ is the average nuclear overlap function
as described in Ref.~\cite{ALICE-PUBLIC-2018-011} and given in Table~\ref{tab:TAA} for the centrality intervals used for the analyses at midrapidity and forward rapidity. The \RAA is evaluated as a function of the average number of participant nucleons, \Npart, corresponding to a given centrality class (as shown in Table~\ref{tab:TAA}), and as a function of the \jpsi \pt. The differential \jpsi production cross section in \pp collisions, $\mathrm{d}^{2} \sigma_{\mathrm{pp}} / (\mathrm{d} y \mathrm{~d} \pt)$, was measured at both midrapidity and forward rapidity as reported in Refs.~\cite{ALICE:2019pid} and~\cite{ALICE:2021qlw}, respectively. At midrapidity, the data sample size does not allow to obtain the cross section for $\pt > 10$~GeV/c and an extrapolation is applied for the last \pt interval (10 $<$ \pt $<$ 15 \GeVc), the result of the extrapolation is 6.82 $\pm$ 0.91 nb. The details of this approach are described in Ref.~\cite{Bossu:2011qe,ALICE:2022zig}, and the corresponding \RAA is shown as the open points in Figs.~\ref{fig:RAA_vs_pt}, ~\ref{fig:RAA_vs_pt_mid_vs_forward_y} and ~\ref{fig:RAA_vs_pt_model}.

The \pt-differential \jpsi yields at midrapidity were studied further by extracting the \meanpt and \meanptsq in fine centrality intervals, as described later in this section. For quantitative comparisons of the \pt distributions, the ratio \rAA of the \jpsi \meanptsq measured in \PbPb collisions to the one obtained in \pp collisions at the same energy is calculated as
\begin{equation}
\rAA=\frac{\meanptsq_{\rm PbPb}}{\meanptsq_{{\rm pp}}}.
\label{eq:little_raa}
\end{equation}

\begin{table}[!htb] 
\begin{center}
\caption{Average nuclear overlap function $\langle$\TAA$\rangle$ and the average number of participants $\langle$\Npart$\rangle$ in \PbPb collisions at $\sqrt{s_{\rm{NN}}}= 5.02 $~TeV for the centrality classes used in the analyses at midrapidity (upper) and forward rapidity (lower).}
\begin{tabular}[t]{c|c|c}
Centrality & $\langle$$T_{\rm{AA}}$$\rangle$ (1/mb) & $\langle$\Npart$\rangle$ \\
\hline
0--5\% &  26.08 $\pm$ 0.18   & 383.40  $\pm$ 0.57 \\
5--10\% &  20.44 $\pm$ 0.17  & 331.20 $\pm$ 1.03 \\
10--20\% &  14.4 $\pm$ 0.13  & 262.00 $\pm$ 1.15 \\
20--30\% &  8.77 $\pm$ 0.10  & 187.90 $\pm$ 1.34  \\
30--40\% &  5.09 $\pm$ 0.08  & 130.80 $\pm$ 1.33\\
40--50\% &  2.75 $\pm$ 0.05  & 87.14 $\pm$  0.93\\
50--70\% &  0.98 $\pm$ 0.02  & 42.65 $\pm$ 0.69 \\
70--90\% &  0.16 $\pm$ 0.004 & 11.34 $\pm$ 0.16 \\
\hline
0--20\% &  18.83 $\pm$ 0.14 & 309.7 $\pm$ 0.89 \\ 
20--40\% & 6.93  $\pm$ 0.09 & 159.4 $\pm$ 1.32\\
40--90\% &  1.00 $\pm$ 0.02 & 39.03 $\pm$ 0.53\\
\hline
\end{tabular}
\label{tab:TAA}
\end{center}
\end{table}

\subsection{ \textbf{J/$\psi$} raw yield extraction}
\label{signal extraction}
The \jpsi mesons are reconstructed employing the \ee decay channel at midrapidity and the \mumu decay channel at forward rapidity. The analysis techniques are discussed in detail in Refs.~\cite{ALICE:2019nrq, ALICE:2015jrl,ALICE:2019lga}. Here, only a brief overview is given and differences with respect to previous analyses are highlighted. 

Electron candidates for the analysis at midrapidity are tracks reconstructed in both the ITS and TPC in the pseudorapidity range \etarange0.9 and with a \pt $>$ 1~\GeVc to suppress combinatorial background.  All tracks are required to have at least one hit in the SPD layers and at least 70 out of a maximum of 159 clusters reconstructed in the TPC. These and other quality criteria that were applied in addition (see Ref.~\cite{ALICE:2019lga}) ensure good tracking resolution and particle identification. They reduce the background electrons from the conversion of photons in the detector material or from long lived weakly-decaying hadrons as well as tracks from pileup collisions, occuring within the readout time of the TPC. 
Electrons and positrons are identified using selections on the specific energy loss, \dEdx, in the TPC gaseous volume. The measured \dEdx is required to be within 3 standard deviations ($\sigma$) relative to the expected electron specific energy loss corresponding to the track momentum, and more than 3.5$\sigma$ different from either the $\pi$ or proton specific energy loss hypotheses. Electrons from photon conversions surviving the track quality criteria are rejected using a technique where candidate electrons are paired with other electrons selected with less strict criteria to enhance the probability of finding the conversion partner, as described in detail in Ref.~\cite{ALICE:2019nrq}.

Muon candidates were selected such that the track pseudorapidity is within the geometrical acceptance of the muon spectrometer, $-4 < \eta < -2.5$, and that the reconstructed track  matches a track segment reconstructed in the trigger system. The transverse position of the tracks at the end of the absorber is required to be $17.6 < R_{\rm abs} < 89.5$~cm in order to reject tracks crossing the thickest part of the absorber. In addition, a selection was applied on the product of the track momentum and the transverse distance to the primary vertex in order to reduce the contamination produced by particles that do not originate from the interaction point.

\begin{figure}[!htb]
\begin{center}
  \includegraphics[width = 7.5 cm]{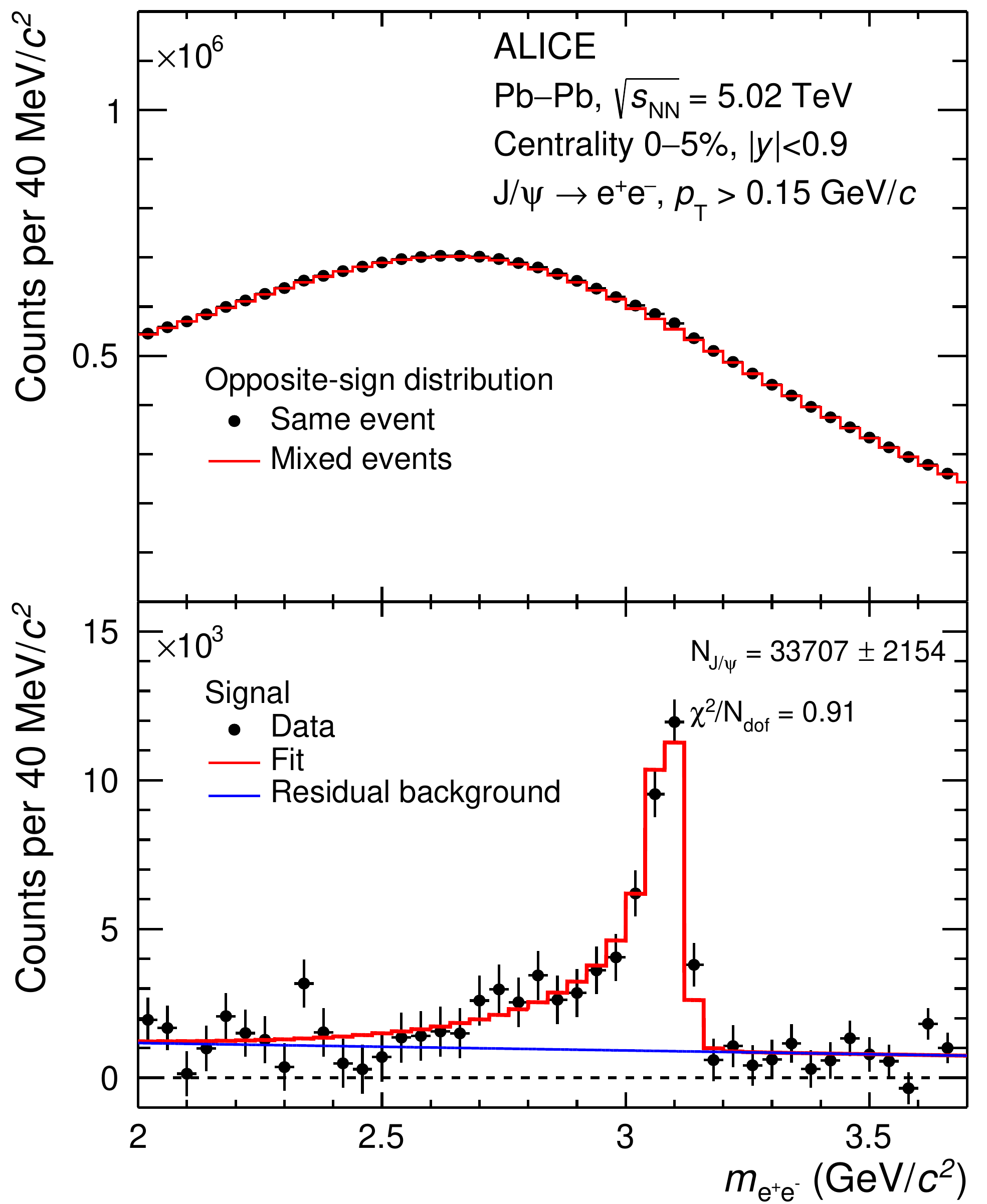}
  \includegraphics[width = 7.5 cm]{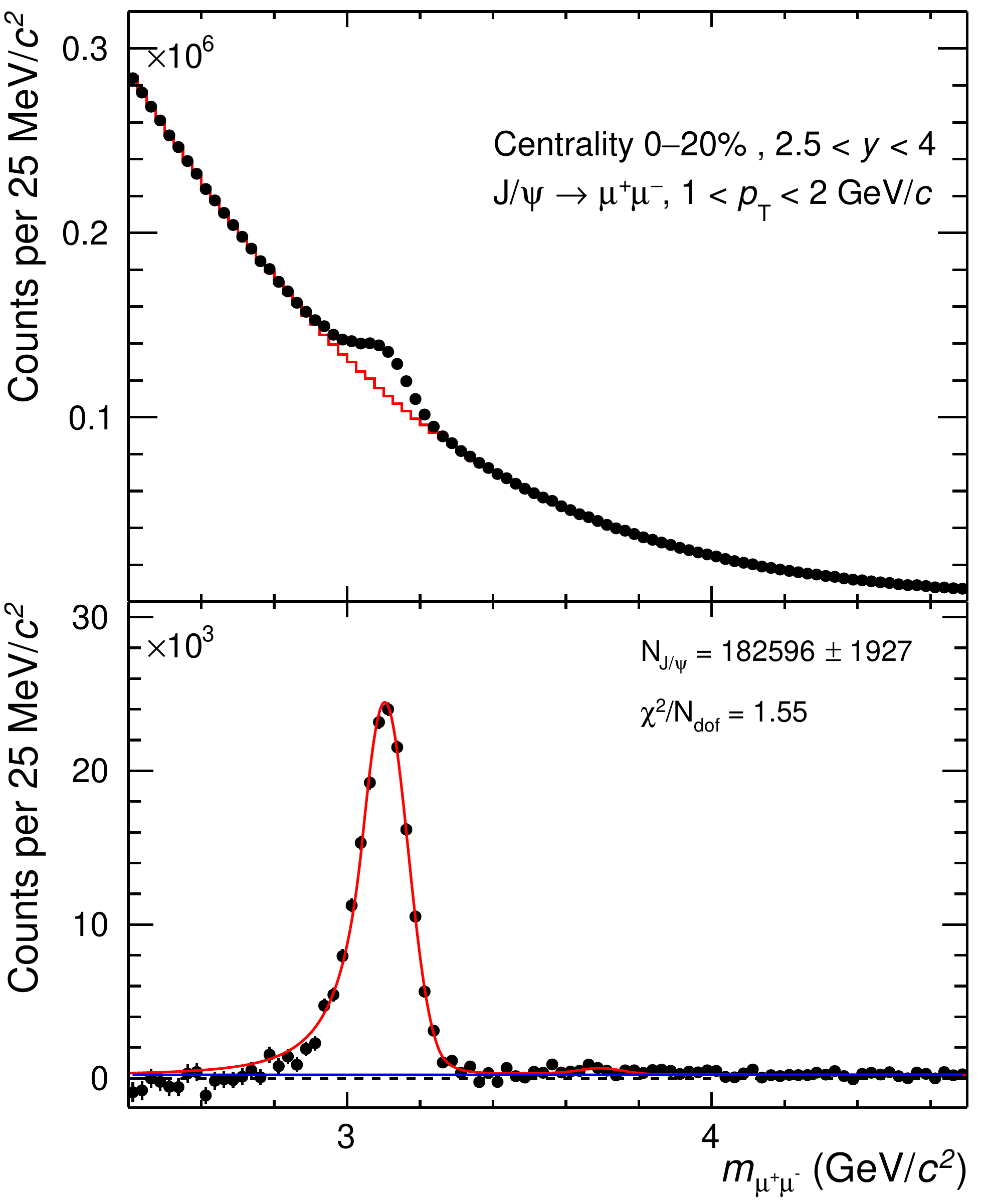}
\caption{ Upper panels: invariant-mass distribution of opposite-sign lepton pairs from the same event (black points) and mixed events (red histograms) at midrapidity (left) and forward rapidity (right) in \PbPb collisions at \fivenn. Lower panels: invariant-mass distribution after the background subtraction with the event-mixing technique. The fit curves, shown in red, represent the sum of the signal and background shapes and the blue curves correspond to the residual background.}
\label{fig:signal_extraction}
\end{center}
\end{figure}

The number of reconstructed \jpsi mesons, i.e. the raw \jpsi yield, was obtained by constructing the invariant-mass distribution of all the possible opposite-sign dileptons with rapidity selections of $2.5 < y < 4$ for dimuons and $|y| < 0.9$ for dielectrons.
At midrapidity, the signal extraction was performed in two steps. First, the combinatorial background was estimated using an event-mixing technique~\cite{ALICE:2019nrq} and subtracted from the invariant-mass distribution. In the second step, the remaining distribution was fitted with a two-component function, one corresponding to the \jpsi signal and the other to the residual background, which mainly arises from correlated semileptonic decays of heavy-flavour hadrons. The \jpsi signal line shape was obtained from Monte Carlo (MC) simulations of \jpsi mesons decaying in the dielectron channel embedded in simulated \PbPb collisions, as described below, while for the residual background a second-order polynomial function was employed. The raw \jpsi yield was obtained by first counting the dielectron pairs in the mass range $2.92 < \mee < 3.16$~\GeVmass\ in the combinatorial-background subtracted invariant-mass distribution, and then subtracting the residual background based on a two-component fit. Finally, the raw \jpsi-meson yield is corrected for the fraction of \jpsi reconstructed outside of the counting mass interval, as described in more detail in Section~\ref{subs_effAcc}. This procedure is illustrated in the left panels of Fig.~\ref{fig:signal_extraction} for the collision centrality interval 0--5\% and $\pt>0.15$~\GeVc. The upper left panel shows the invariant-mass distributions for the opposite-sign dielectrons constructed from the same event (black) and mixed events (red). The fitting procedure of the combinatorial background subtracted invariant-mass distribution discussed above is illustrated in the lower left panel.

At forward rapidity, two different methods were used to extract the number of \jpsi counts. In the first method, the invariant-mass distributions were fitted with a sum of a signal and a background function, while in the second method the event-mixing technique 
was employed, as described in Ref.~\cite{ALICE:2015jrl}.
The fit functions corresponding to the signal are either a double-sided Crystal Ball function (CB2) or a pseudo-Gaussian with a mass-dependent width~\cite{ALICE-PUBLIC-2015-006}. In both cases, the \jpsi pole mass and width were free parameters of the fit, while the non-Gaussian tail parameters were fixed. Two sets of tail parameters were obtained, one based on MC simulations
and one extracted from a large data sample of \pp collisions at \s = 13 TeV~\cite{ALICE:2017leg}. The MC simulations were embedded into real MB events in order to properly account for the effect of the detector occupancy. The \psitwos resonance was also included in the fit to the invariant-mass spectrum, using the same signal function as for the \jpsi with mass and width bound to those of the \jpsi~\cite{ALICE:2017leg,ALICE:2022jeh}. 
The background functions employed in the first method were either a variable-width Gaussian~\cite{ALICE-PUBLIC-2015-006} or a ratio of a second order to a third order polynomial. The residual background in the second method was parameterised with a sum of two exponential functions.  
Finally, two invariant-mass ranges were considered for the fit procedure: $2.2 < m_{\mu^{+}\mu^{-}} < 4.5$ and $2.4 < m_{\mu^{+}\mu^{-}} < 4.7$ \GeVmass. 
An example of the signal extraction fit is shown in the right panel of Fig.~\ref{fig:signal_extraction} before (upper plot) and after (lower plot) the subtraction of the combinatorial background estimated with the event-mixing technique.
For each \pt and centrality interval, several fits were performed with the two different approaches, different combinations of signal and background functions, signal tail parameters, and fitting ranges. The number of \jpsi was obtained as the average of the results from the various fitting methods~\cite{ALICE:2021qlw}. 
These various fitting methods are used to determine the systematic uncertainties on the yield extraction as described in Section~\ref{sec:systematics}. 

About 9.0$\times$10$^5$ and 8.2$\times$$10^4$ raw \jpsi counts are measured at forward rapidity and midrapidity, respectively, integrated over all available centrality and \pt intervals. 
\FloatBarrier

\subsection{ \textbf{J/$\psi$} \meanpt and \meanptsq extraction}\label{jpsiMeanpT}
\begin{figure}[!htb]
\begin{center}
  \includegraphics[width = 7.5 cm]{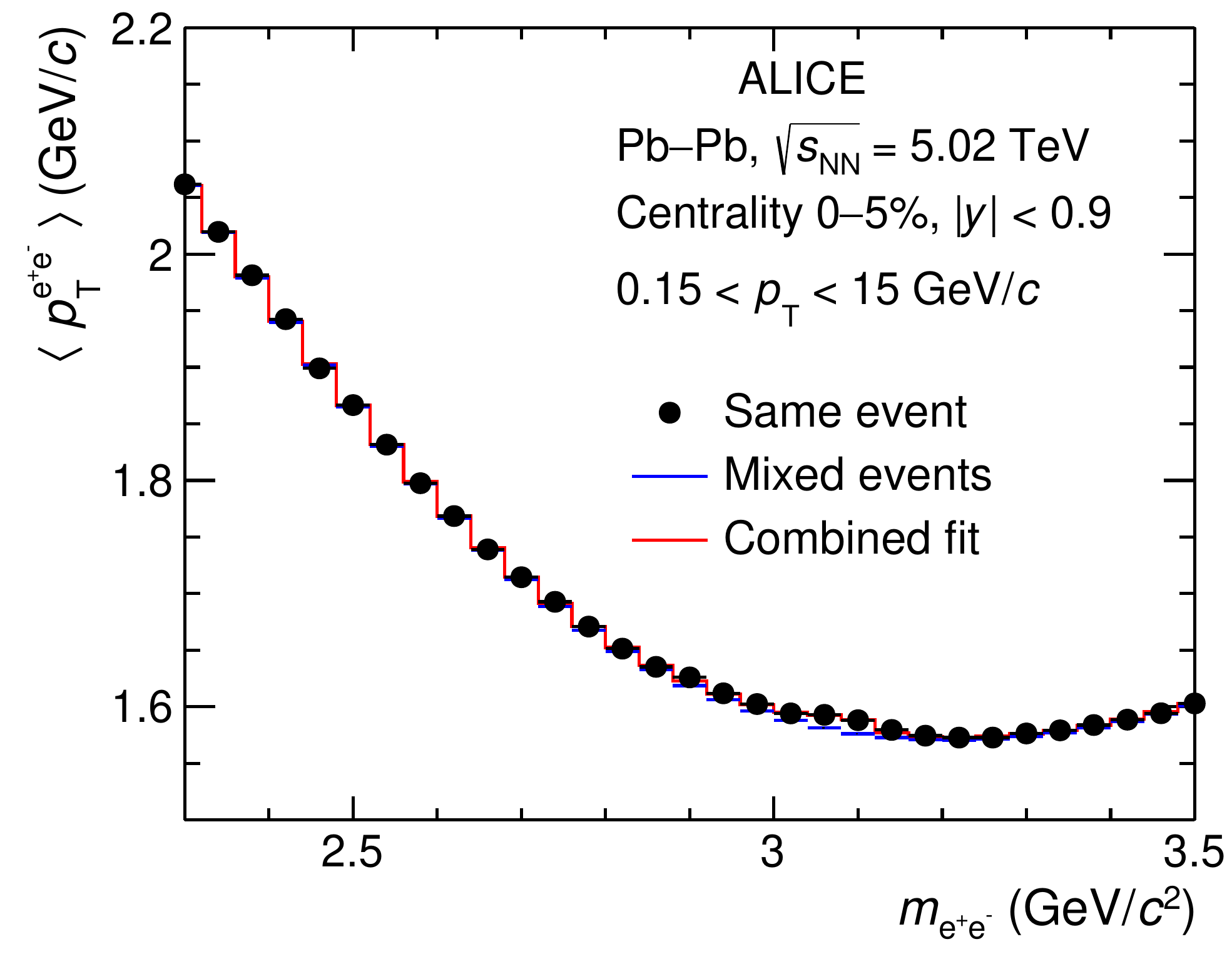}
  \includegraphics[width = 7.5 cm]{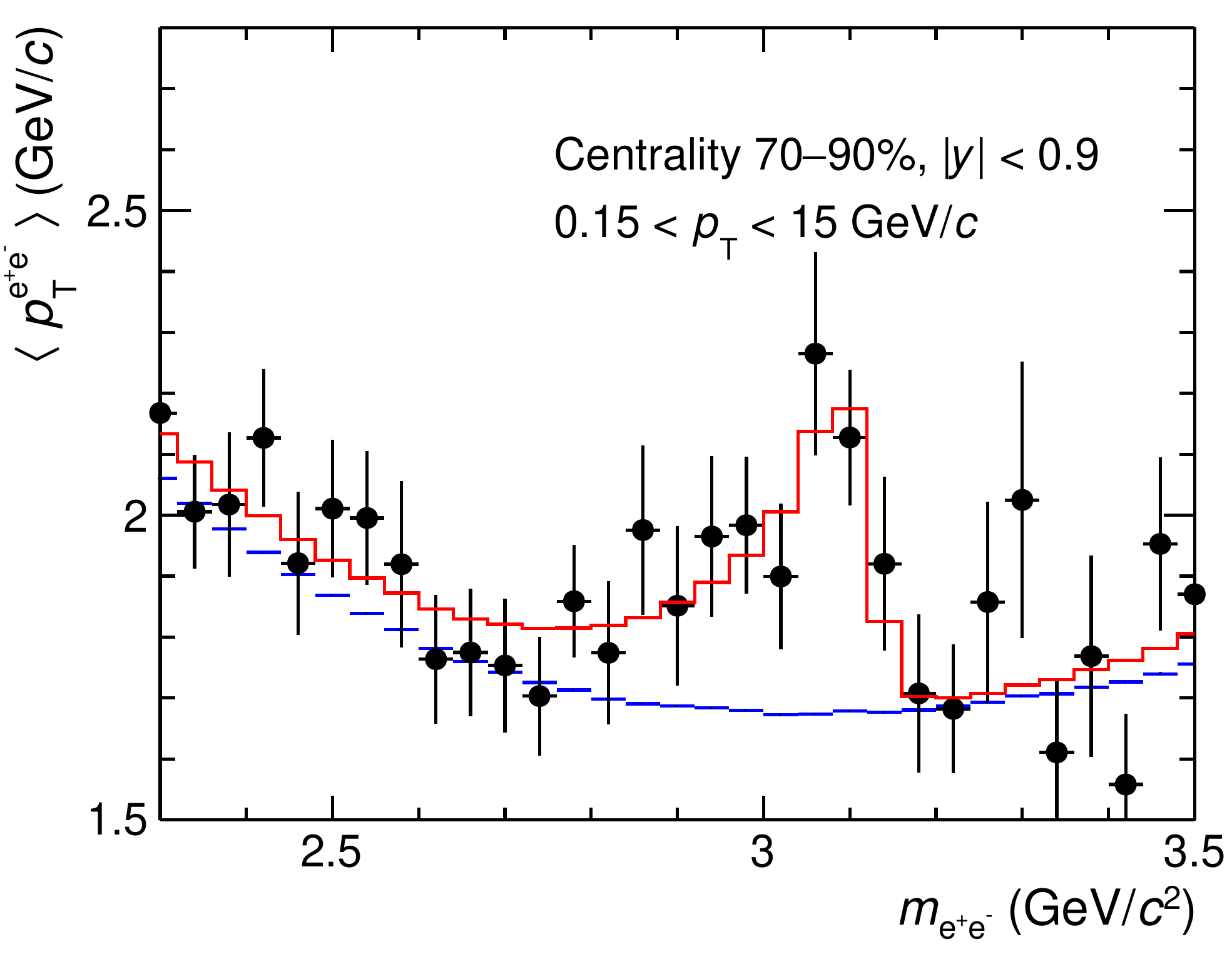}
\caption{\jpsi \meanpt extraction  in \PbPb collisions at \fivenn at midrapidity for the 0--5\% (left panel) and the 70--90\% (right panel) centrality interval. The data points correspond to opposite-sign \ee pairs from the same event, the blue line to the \ee pairs from mixed events, and the red line is the combined fit that includes the mixed events and residual background which is described by the polynomial function.} 
\label{fig:meanpt}
\end{center}
\end{figure}

At midrapidity, a quantitative study of the \jpsi \pt spectrum in fine centrality intervals is conducted by extracting the \jpsi mean \pt, $\meanpt_{\jpsi}$, and mean \pt squared, $\meanptsq_{\jpsi}$. For a given centrality interval, these quantities are obtained based on a fit to the mass dependent \meanpt and \meanptsq distributions after efficiency correction, using a function defined as:

\begin{equation}
X (\mee) = f(\mee) \times X_{\jpsi} + (1-f(\mee)) \times X_{\mathrm{bkg}}(\mee)
\label{eq:jpsimeanpt}
\end{equation}

where $X$ stands for either the \meanpt or \meanptsq, and $f$ is the invariant-mass dependent fraction of \jpsi signal determined in the signal-extraction procedure explained above. The invariant-mass dependent background component $X_{\mathrm{bkg(m_{e^{+}e^{-})}}}$ is determined from the event-mixing procedure plus a second order polynomial function for the residual background. Examples of these fits for the \meanpt observable are shown in Fig.~\ref{fig:meanpt}.

\FloatBarrier

\subsection{Acceptance and efficiency correction}\label{subs_effAcc}

The acceptance times reconstruction efficiency factor (\AccEff), which enters into Eq.~\ref{eq:d2ndydpt}, was computed employing both MC and data-driven methods. At midrapidity, this factor includes the kinematic acceptance, track-reconstruction and particle-identification efficiencies, and the fraction of \jpsi with an invariant mass in the signal counting range. With the exception of the particle identification, obtained with a data-driven method, the corrections were obtained using a MC simulation of \jpsi embedded in simulated \PbPb collisions. The \PbPb collisions were generated using the HIJING 1.0 model~\cite{Wang:1991hta}, while the \jpsi were generated using a cocktail of prompt \jpsi with a kinematic distribution tuned to existing measurements and non-prompt \jpsi from beauty hadrons forced to decay into channels containing \jpsi, using PYTHIA 6.4~\cite{pythia}. The \ee decay of the embedded \jpsi was handled using PHOTOS ~\cite{photos}. Both the prompt \jpsi and the beauty hadrons forced to decay into non-prompt \jpsi were assumed to be unpolarised, in agreement with existing measurements, which indicate small or no polarisation~\cite{acharya_measurement_2018,ALICE:2020iev}. All generated particles were transported through the ALICE detector setup using GEANT3~\cite{geant3}, taking into account the time dependence of detector conditions during the 2015 and 2018 data-taking periods. For the determination of the particle-identification efficiency, a clean sample of electrons from photon conversions, passing similar quality selection criteria as primary electrons in the TPC, is used to compute differential maps in pseudorapidity, azimuthal angle $\varphi$ and momentum $p$ for the single-electron selection efficiency. These were then propagated to the \jpsi dielectron pairs using the phase-space distribution of the \jpsi decay simulated in the above mentioned MC simulations.
The total ($A\times \varepsilon$) for the \pt-integrated \jpsi yields is about 6.5\% in the 0--10\% centrality interval, and it slightly increases towards more peripheral collisions. As a function of \pt, the ($A \times \varepsilon$) has a non-monotonic behaviour, with a minimum value of 5.6\% around \pt=~2~\GeVc, and a maximum of about 9\% towards zero and high \pt.

At forward rapidity, the acceptance and reconstruction efficiency values were determined using simulated \jpsi mesons forced to decay via the dimuon channel, embedded into real events. The \jpsi \pt- and \y-differential distributions used in the simulation were adjusted to measurements via an iterative procedure, and separately for all centrality intervals employed in this analysis. 
The \jpsi were assumed to be unpolarised, in agreement with the small polarisation, compatible with zero, measured in \PbPb collisions for $2<\pt<10$~\GeVc~\cite{ALICE:2020iev}. As in the analysis at midrapidity, the simulations take into account the time dependence of the detector conditions, such as the status of the tracking chambers and the residual detector element misalignment. The trigger-chamber efficiency was determined from data and used as input in the simulations. The (\AccEff) reaches a minimum of 11\% at \pt $\approx$~2~GeV/c and increases up to 13\% at low \pt and up to 46\% at high \pt in the 0--20\% centrality interval. It increases towards peripheral collisions by a few percent~\cite{Abelev:2014ffa}.

\FloatBarrier

\subsection{Systematic uncertainties}
\label{sec:systematics}
The considered sources of systematic uncertainty for the analysis at midrapidity include central-barrel tracking, electron identification, signal extraction, and the kinematics of the \jpsi injected in the MC simulations. For the analysis at forward rapidity, the main systematic uncertainties originate from the signal extraction, the muon tracking and trigger efficiencies, and the kinematics of the \jpsi used in the embedded MC simulations. In addition, for the $\RAA$, uncertainties on the pp reference and the nuclear overlap function are included in both analyses ~\cite{ALICE-PUBLIC-2018-011}. The uncertainty on the \jpsi decay branching ratio and the evaluated systematic uncertainties are summarised in Table~\ref{table:systematics} and Table~\ref{tab:muon_syst} for the analysis at midrapidity and forward rapidity, respectively. 

At midrapidity, the tracking uncertainty is the largest source of systematic uncertainty and it is dominated by the ITS--TPC track matching. This was determined based on the difference in the matching efficiency observed for single tracks between data and MC simulations. The systematic uncertainty due to the electron identification takes into account the residual miscalibration of the TPC particle identification (PID) response and also the statistical uncertainty of the clean electron sample used to compute the identification efficiency. For its estimation, the PID selection criteria (both electron inclusion and hadron rejection) are varied, each time obtaining a new set of raw yields and corresponding PID efficiencies. The assigned systematic uncertainty is taken as the standard deviation of the distribution of the corrected results obtained in this procedure. The systematic uncertainty of the signal extraction includes one component from the \jpsi signal shape obtained from MC simulations and one component related to the description of the dielectron background. The former was determined by varying the mass range in which the signal is counted and recomputing each time the corresponding signal fraction correction, while the latter was determined by repeating the fit to the invariant-mass distribution in different mass ranges. The standard deviation of the corrected yield distribution was then taken as the systematic uncertainty.
Since the \jpsi efficiency depends on \pt, the average efficiency computed over wide \pt intervals depends in turn on the underlying \jpsi \pt distribution used in the MC simulations. The corresponding uncertainty was minimised by iteratively tuning the injected \pt spectrum to match the corrected spectrum measured in this analysis. A systematic uncertainty which takes into account possible variations of the \pt spectrum, statistically compatible with the finally measured \pt spectrum, was assigned and it is typically below 1\%. The total systematic uncertainty on the \jpsi corrected yield varies in the range 6--10\% in different \pt and centrality intervals. The systematic uncertainties, which are dominated by the tracking uncertainties, are correlated over centrality and \pt to a very large extent. 
The systematic uncertainties for \meanpt and \meanptsq are evaluated via similar procedures as for the \pt integrated yields. The uncertainties from signal extraction and track selection criteria range, respectively, from 0.2 to 1.2\% and from 0.5 to 1.3\%. The electron identification and ITS--TPC matching systematic uncertainties are calculated by propagating the \pt-differential systematic uncertainties to the \meanpt based on the measured \pt-differential spectrum.

In the analysis at forward rapidity, the systematic uncertainty corresponding to the signal extraction was determined using several variations of the fit to the invariant-mass spectra, including the fit method, the signal and background functions, and the fitted mass range. 
This uncertainty varies in the range 1.5--10.7\% depending on the \pt interval and centrality class. 
The systematic uncertainty on (\AccEff) depends on the uncertainty on the \pt and \rapidity distributions of the simulated \jpsi, and on the tracking, trigger, and matching efficiency. The first two were evaluated by varying the \pt and \rapidity spectrum for each centrality interval, taking into account the correlations between the \pt and \rapidity distributions. The systematic uncertainty was estimated as the largest difference between the nominal (\AccEff) and the one estimated from the variations. It ranges between 0.2\% and 4.1\%. The systematic uncertainty on the muon tracking efficiency was estimated based on the difference between the single-muon tracking efficiency obtained in data and MC with a method that uses the redundancy of the tracking information in each station. The corresponding uncertainty for dimuons was evaluated to be 3\% and constant over \pt. An additional systematic uncertainty
is ascribed to the loss of tracking efficiency due to occupancy effects in the most central events and was estimated to range between 0.5 and 1\%, increasing towards more central events. 
The systematic uncertainty on the trigger efficiency has two components, one due to the intrinsic efficiency of the trigger chambers and another one due to the trigger response. The first component was estimated from the uncertainties on the single-muon trigger efficiency measured from data and used in the simulations.  
The second component was evaluated by comparing the \pt dependence of the trigger response function of the single muon between data and MC. 
The two sources were added in quadrature and the obtained uncertainty ranges between 1.5 and 2\%. Finally, a 1\% systematic uncertainty is assigned, related to the choice of the $\chi^2$ selection used to define the matching between the tracks reconstructed in the tracking system and the track segments reconstructed in the trigger system.
The uncertainty on $F_{\mathrm{norm}}$ was estimated by using two methods. First, the opposite-sign dimuon trigger condition was applied when analysing recorded MB events and, second, the counting rate of the dimuon and MB triggers were compared. The estimated uncertainty on $F_{\mathrm{norm}}$ was obtained by comparing the two methods and it amounts to 0.7\%. 

The systematic uncertainties on $\langle$\TAA$\rangle$ were obtained as described in Ref.~\cite{ALICE-PUBLIC-2018-011} and the values are listed in Table~\ref{tab:TAA} for both analyses, at midrapidity and forward rapidity. 
The systematic uncertainty on the definition of the centrality interval was estimated using variations of $\pm 1\%$ of the V0 signal amplitude corresponding to 90\% of the hadronic Pb--Pb cross section and redefining correspondingly the centrality intervals. The systematic uncertainty of the centrality limit depends on the width of the centrality classes and it ranges from 1 to 6\% and from 0 to 2.8\%, as shown in Tables ~\ref{table:systematics} and ~\ref{tab:muon_syst} for the analyses at midrapidity and forward rapidity, respectively. 

The systematic uncertainties on the \jpsi reference cross section in pp collisions at \five as obtained in Refs.~\cite{Acharya:2019lkw, ALICE:2021qlw} are provided in Table~\ref{table:systematics} and Table~\ref{tab:muon_syst}. 
The correlations of the systematic uncertainties over centrality and \pt depend on the mid and forward rapidity analysis and they are indicated in Table~\ref{table:systematics} and ~\ref{tab:muon_syst}.

    \begin{table}[!htb]
    	\begin{center}
    	\caption{Systematic uncertainties on the $\pt$-integrated ($0.15<p_\textrm{T}<15~ \textrm{GeV/}c$) measurement at midrapidity for different centrality intervals. The individual contributions and the total uncertainties are given in percentage. It is considered that all the uncertainties are correlated over centrality and \pt to a very large extent.}
    	\label{table:systematics}
    	\begin{tabular}{c | cccccccc} 
    		\hline
    		Centrality (\%)&0--5 &5--10 &10--20 &20--30 &30--40& 40--50 &50--70 &70--90\\
    		\hline
    		Signal extraction & 2.5 & 2.7 & 1.9 & 6.6 & 2.5 & 1.0 & 1.8 & 2.5\\
            MC input & 0.5 & 0.5 & 0.5 & 0.5  & 0.5  & 0.5 & 0.5 & 0.5 \\
    		Tracking & 10.1 & 10.1 & 8.5 & 8.5 & 8.0 & 8.0 & 7.9 & 7.9\\
    		PID & 1.6 & 1.3 & 1.4 & 1.4 & 1.1 & 1.2 & 1.1 & 1.3\\
    		Centrality limit & 1.0 & 1.0 & 1.0 & 1.0  & 1.0  & 1.0 & 5.7 & 6.0 \\
    		\hline
    		Total & 10.6 & 10.5 & 8.8 & 10.8 & 9.0 & 8.2 & 10.0 & 10.3\\ 
    		\hline
    		\TAA (only on \RAA) & 0.7 & 0.8 & 0.9 & 1.2  & 1.6  & 1.7 & 2.0 & 2.3 \\
    		pp reference (only on \RAA) & \multicolumn{8}{c}{5.8} \\
    		
    		Branching ratio (only on yield) &  \multicolumn{8}{c}{0.5} \\
            \hline
    	\end{tabular}
    	\end{center}
    \end{table}

    \begin{table}[htbp] 
    \caption{
    Systematic uncertainties on the $\pt$-differential measurement at forward rapidity for various centrality intervals. The individual contributions are given in percentage. When a range is given, it  corresponds to the minimum and maximum values obtained in the \pt interval. Values marked with an asterisk correspond to the uncertainties correlated over \pt. These uncertainties are considered as global ones for \RAA and added in quadrature to the uncorrelated uncertainties for the yields.}
\begin{center}
\vspace{1ex}
\begin{tabular}[t]{c|ccc}
\hline
Centrality (\%) & 0--20\% & 20--40\% & 40--90\% \\  
\hline
$F_{\rm{norm}}$ & \multicolumn{3}{c}{0.7*} \\ 
\hline 
Signal extraction & 1.5--5.8 & 1.9--4.5 & 1.6--10.7\\
MC input & 1.8--4.1 & 0.2--2.1 & 0.9--1.8\\
Tracking efficiency & 3.0 + 1.0* & 3.0 + 0.5* & 3.0\\
Trigger efficiency & 1.5--2.0 + 1.0* & 1.5--2.0 + 0.5* & 1.5--2.0\\
Matching efficiency & 1.0 & 1.0 & 1.0 \\ 
Centrality limit & -- & 0.8* & 2.8*\\
\hline
$T_{\rm{AA}}$ (only on \RAA) & 0.8* & 1.3* & 2.0*\\
pp reference (only on \RAA) & \multicolumn{3}{c}{3.5--5.6 + 1.9*}\\
Branching ratio (only on yield)&  \multicolumn{3}{c}{0.5*} \\
\hline 

\end{tabular}
\end{center}
\label{tab:muon_syst}
\end{table}

\FloatBarrier

\section{Results}\label{sec:results} 

\subsection{Inclusive \textbf{J/$\psi$} yields} \label{subsec:spectrum}

\begin{figure}[!htb]
\begin{center}
  \includegraphics[width = 7.5 cm]{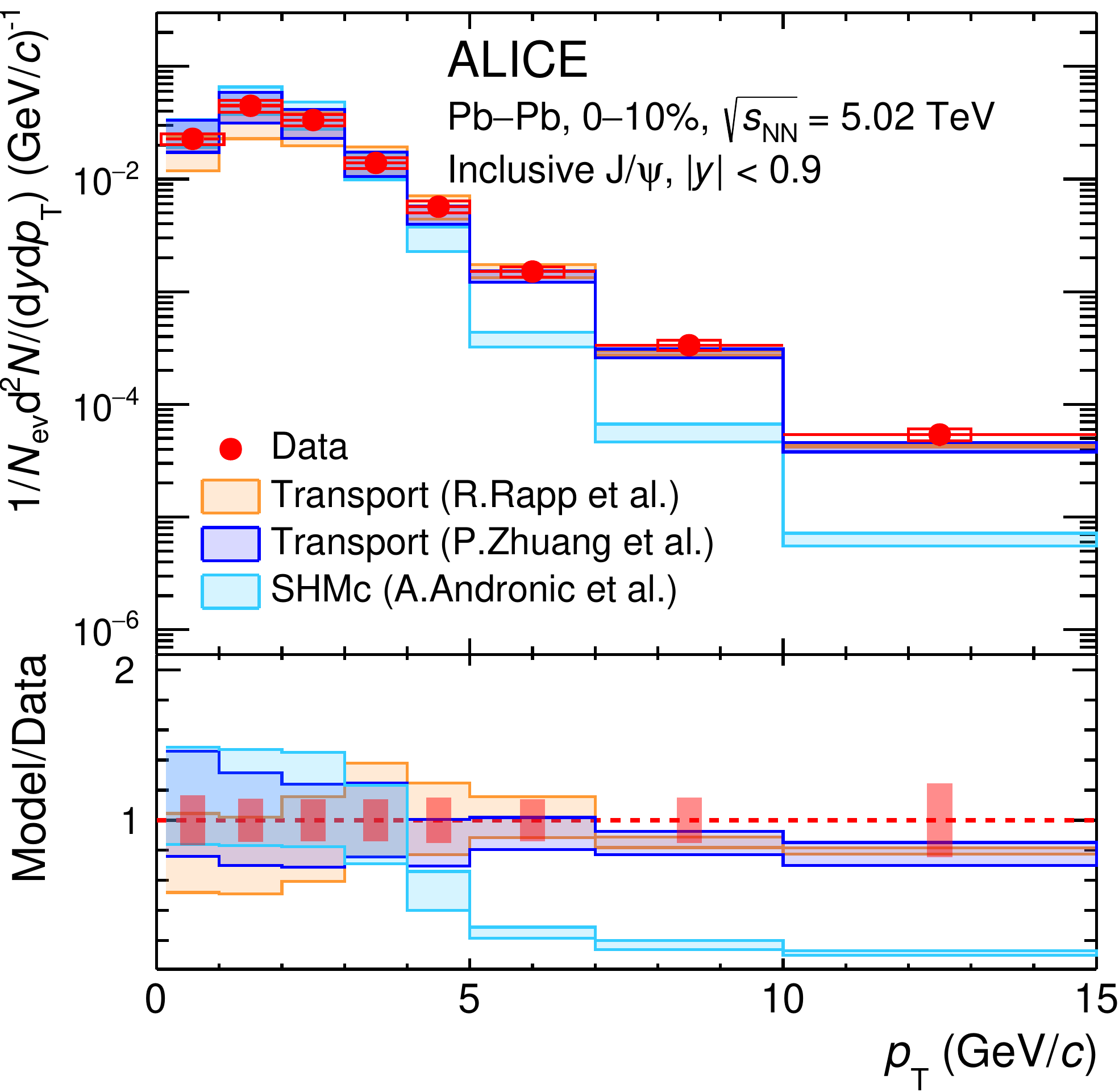}
  \includegraphics[width = 7.5 cm]{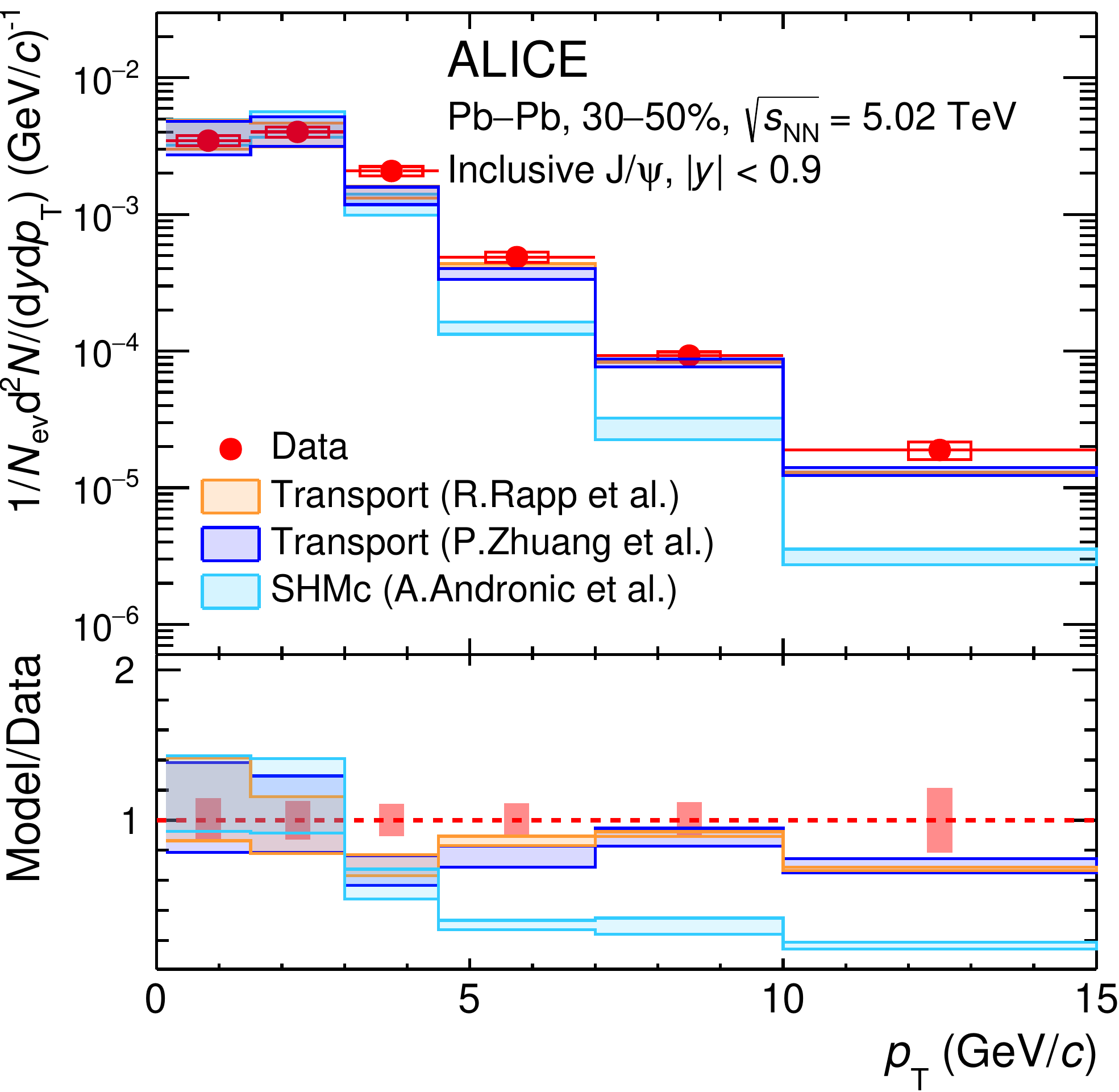}
\caption{\jpsi \pt-differential production yields in $\PbPb$ collisions at \fivenn at midrapidity in the $0$--$10$\% (left panel) and $30$--$50$\% (right panel) centrality intervals. The statistical and systematic uncertainties are indicated, respectively, by the vertical error bars and the open boxes. The horizontal bars indicate the \pt intervals. Data are compared to model calculations from Refs.~\cite{Andronic:2019wva,Zhao:2007hh,Zhou:2014kka}. The ratios between data and models are shown in the lower panels. The filled boxes around unity depict the quadratic sum of statistical and systematic uncertainties from the measurement, while the bands indicate model uncertainties.}
\label{fig:JpsiPtSpectra_mid}
\end{center}
\end{figure}

\begin{figure}[!htb]
\begin{center}
  \includegraphics[width = 7.5 cm]{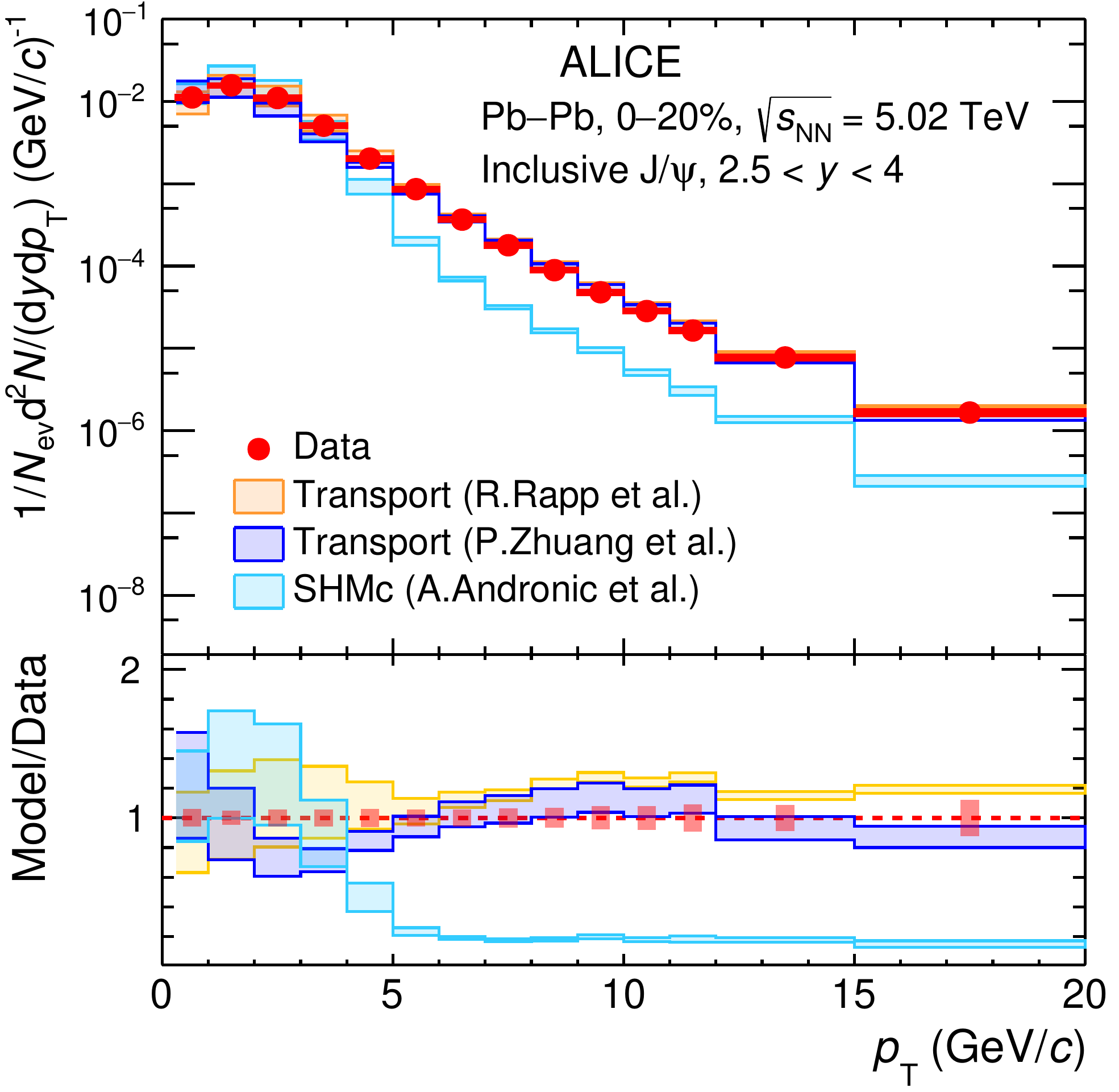}
  \includegraphics[width = 7.5 cm]{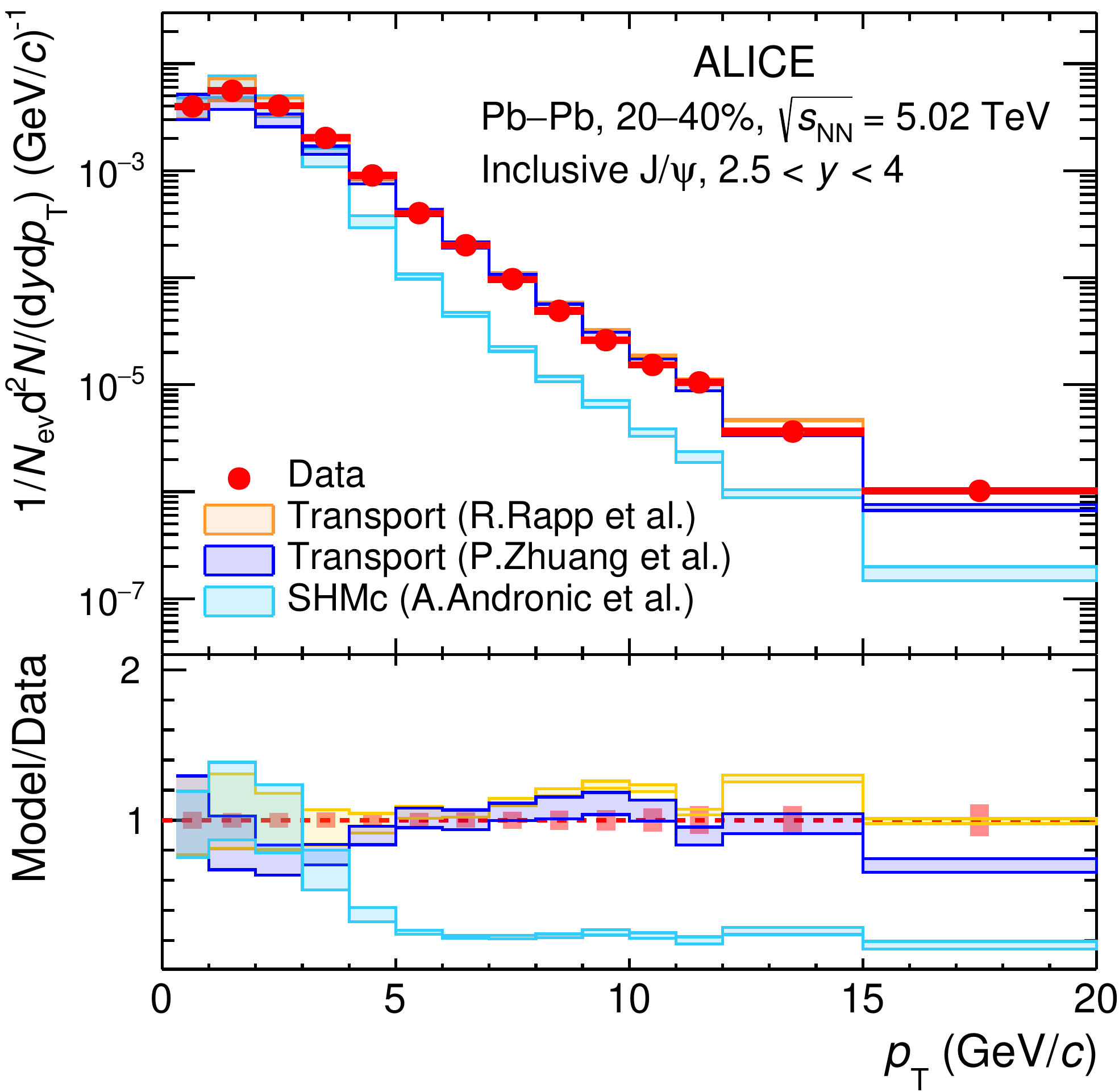}
  \includegraphics[width = 7.5 cm]{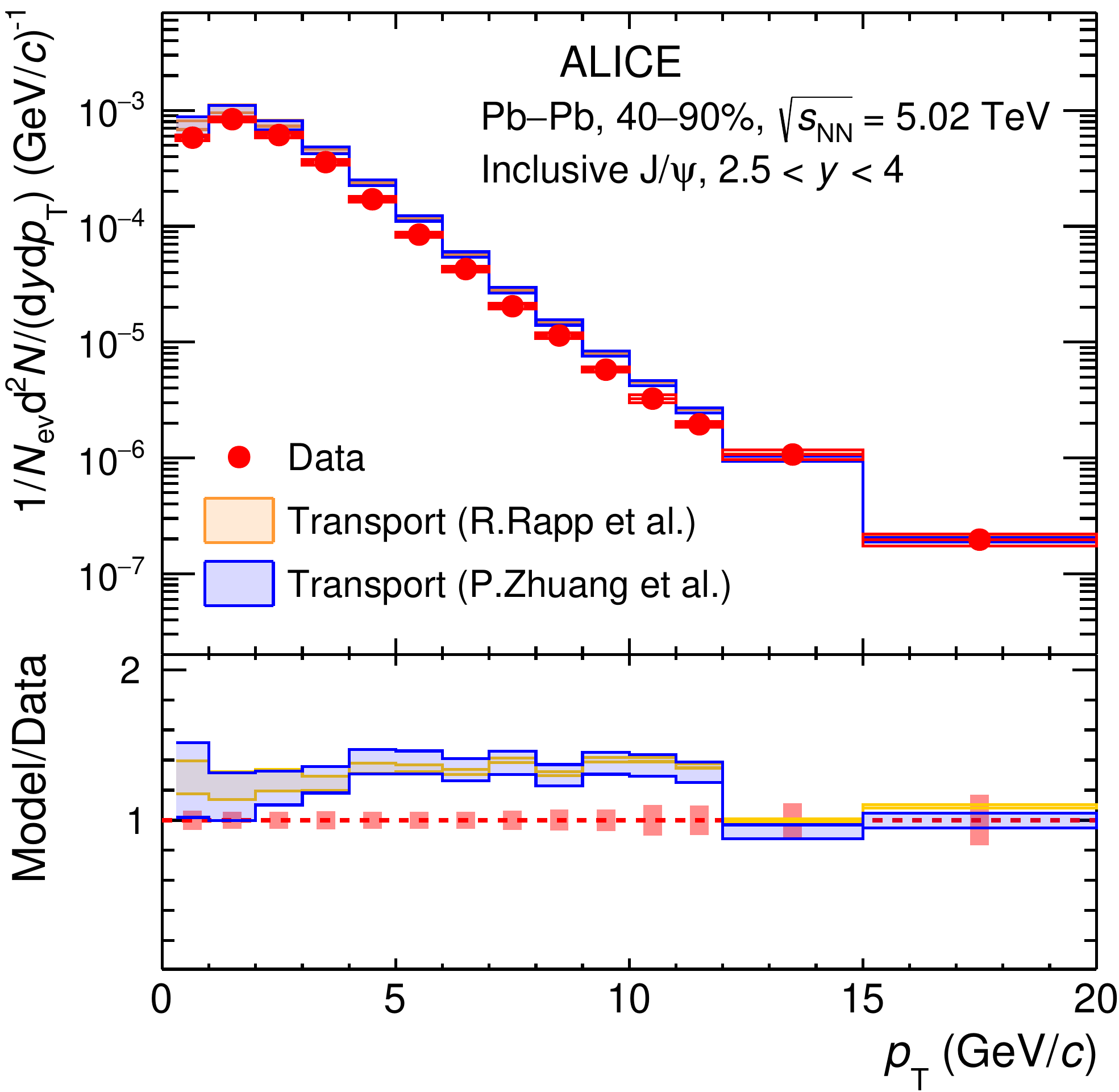}
\caption{\jpsi \pt-differential production yields in $\PbPb$ collisions at \fivenn at forward rapidity in the $0$--$20$\%, $20$--$40$\%, and $40$--$90$\% centrality intervals. The statistical and systematic uncertainties are indicated, respectively, by the vertical error bars and the open boxes. The horizontal bars indicate the \pt intervals. Data are compared to model calculations from Refs.~\cite{Andronic:2019wva,Zhao:2007hh,Zhou:2014kka}. The ratio between data and models is shown in the lower panels. The filled boxes around unity depict the quadratic sum of statistical and systematic uncertainties from the measurement, while the bands indicate model uncertainties.} 
\label{fig:JpsiPtSpectra_forward}
\end{center}
\end{figure}

The fully corrected inclusive \jpsi \pt-differential yields, \dndydpt, were obtained according to Eq.\ref{eq:d2ndydpt} in centrality intervals in $\PbPb$ collisions at \fivenn at midrapidity (\yrange{0.9}) and at forward rapidity ($2.5 < \y < 4$). Figure~\ref{fig:JpsiPtSpectra_mid} shows the \jpsi yields obtained at midrapidity for the $0$--$10$\% and $30$--$50$\% centrality intervals, while Fig.~\ref{fig:JpsiPtSpectra_forward} shows the yields measured at forward rapidity in the $0$--$20$\%, $20$--$40$\% and $40$--$90$\% centrality intervals. For all of the results, the statistical and systematic uncertainties are indicated by the vertical error bars and the open boxes around the data points, respectively. These results are compared with calculations performed using  the statistical hadronisation model (SHMc) by Andronic et al.~\cite{Andronic:2019wva}, and two microscopic transport models by Rapp et al.~\cite{Zhao:2007hh} and Zhuang et al.~\cite{Zhou:2014kka}. The physics assumptions in these model calculations are discussed in more detail in the next section. The lower panels of the Figs.~\ref{fig:JpsiPtSpectra_mid} and ~\ref{fig:JpsiPtSpectra_forward} depict the ratio between the experimental data and the different model calculations, with the width of the bands representing the model uncertainties. These uncertainties are due to uncertainties on input parameters, mainly the total charm-quark production cross section and CNM effects. The filled boxes around unity represent the uncertainties of the measured results, shown as the quadratic sum of statistical and systematic uncertainties. Both transport models describe the \pt-differential yields in central and semicentral collisions, while they overestimate them at forward rapidity in peripheral collisions. The SHMc calculations coupled with a hydrodynamics inspired freeze-out parameterisation are in good agreement with the data in the low-$\pt$ region, but underestimate the measurements at higher \pt.

\FloatBarrier

\subsection{The \textbf{J/$\psi$} nuclear modification factor \RAA } \label{subsec:RAA}

\begin{figure}[!htb]
\begin{center}
  \includegraphics[width = 7.5 cm]{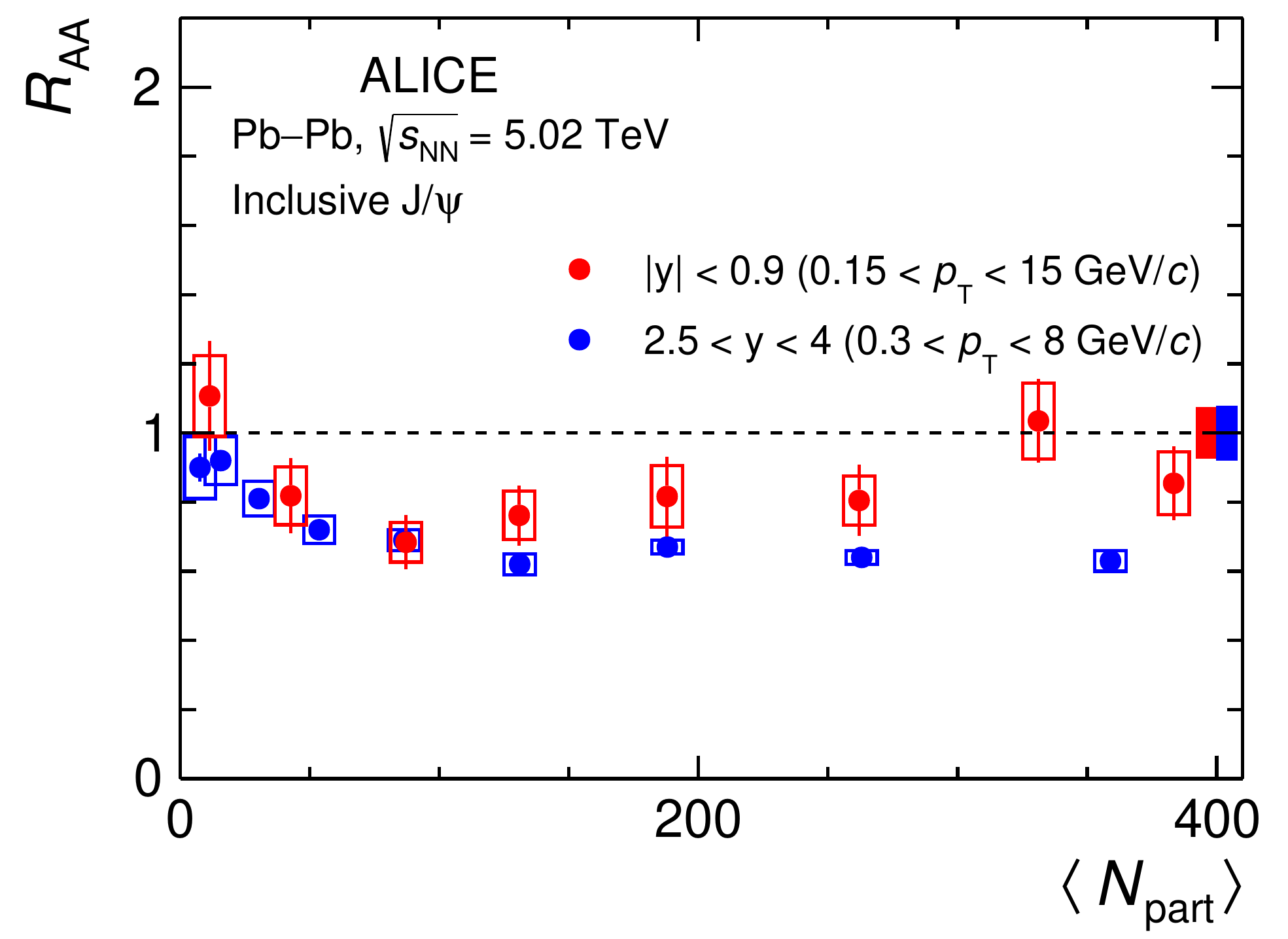}
\caption{Inclusive \jpsi nuclear modification factor at midrapidity and forward rapidity~\cite{ALICE:2016flj}, integrated over \pt, as a function of the number of participants in \PbPb collisions at \fivenn. The statistical and systematic uncertainties are indicated, respectively, by the vertical error bars and the open boxes around the data points. The filled boxes around unity show the global uncertainties.}
\label{fig:Raa_vs_cent}
\end{center}
\end{figure}

The nuclear modification factor \RAA was obtained using the measured yields, according to Eq.~\ref{eq:Raa}. Figure~\ref{fig:Raa_vs_cent} shows the \pt-integrated \jpsi \RAA as a function of \Npart in \PbPb collisions at \fivenn, obtained in the current analysis at midrapidity in comparison to the results at forward rapidity, previously reported by the ALICE Collaboration in Ref.~\cite{ALICE:2016flj}. Global uncertainties represented the centrality-correlated uncertainties, shown as filled boxes around unity, and are largely uncorrelated between the forward and midrapidity analyses. Both results exclude low-\pt \jpsi, with a selection of $\pt>0.15$~\GeVc and $\pt>0.3$~\GeVc at midrapidity and forward rapidity, respectively, in order to reject \jpsi produced via photoproduction processes~\cite{ALICE:2012yye,ALICE:2019tqa,ALICE:2021gpt}, which contribute significantly to the \jpsi yield in particular in peripheral collisions~\cite{ALICE:2015mzu,ALICE:2022zso}. The \RAA is compatible with unity in the most peripheral collisions, while a suppression of the \jpsi production in \PbPb collisions with respect to binary scaled \pp collisions is observed in semicentral and central collisions, in particular at forward rapidity. At midrapidity, \RAA exhibits a slightly increasing trend from approximately $\langle\Npart\rangle$~=~100 towards the most central collisions, with slightly larger \RAA values at midrapidity than at forward rapidity, which confirms previous observations reported by ALICE~\cite{ALICE:2019nrq}. The results at midrapidity are larger than those measured at forward rapidity, with a significance of the difference of $2.2\sigma$ when considering the data points in the centrality range 0--10\%. 
The larger \RAA in central collisions at midrapidity is expected in phenomenological models due to the larger ${\rm d}\sigma_{\ccbar}/{\rm d}y$ at midrapidity which leads to larger fraction of \jpsi produced via (re)generation~\cite{Andronic:2019wva,Zhou:2014kka,Wu:2020zbx}.

\begin{figure}[!htb]
\begin{center}
  \includegraphics[width = 7.5 cm]{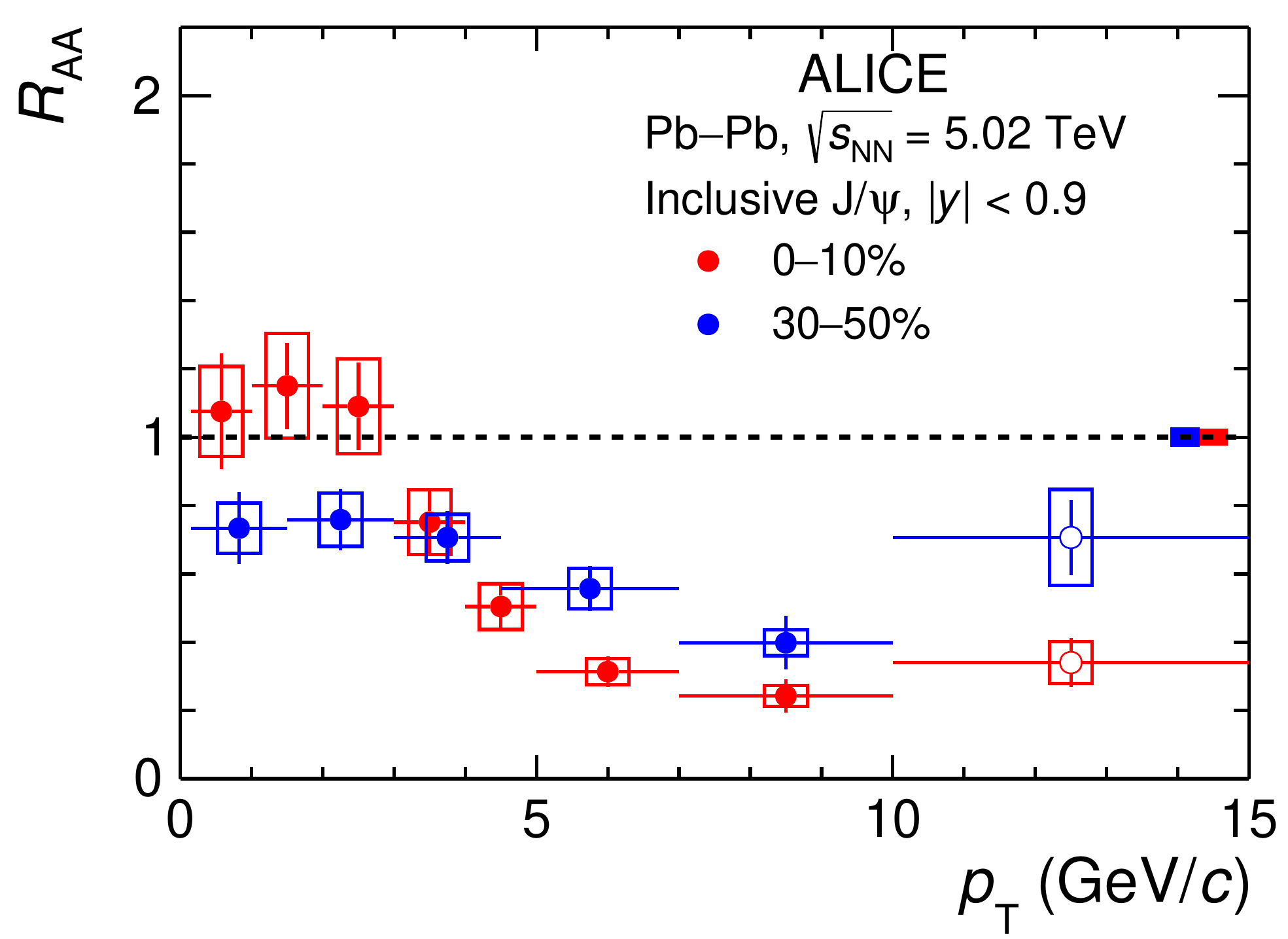}
  \includegraphics[width = 7.5 cm]{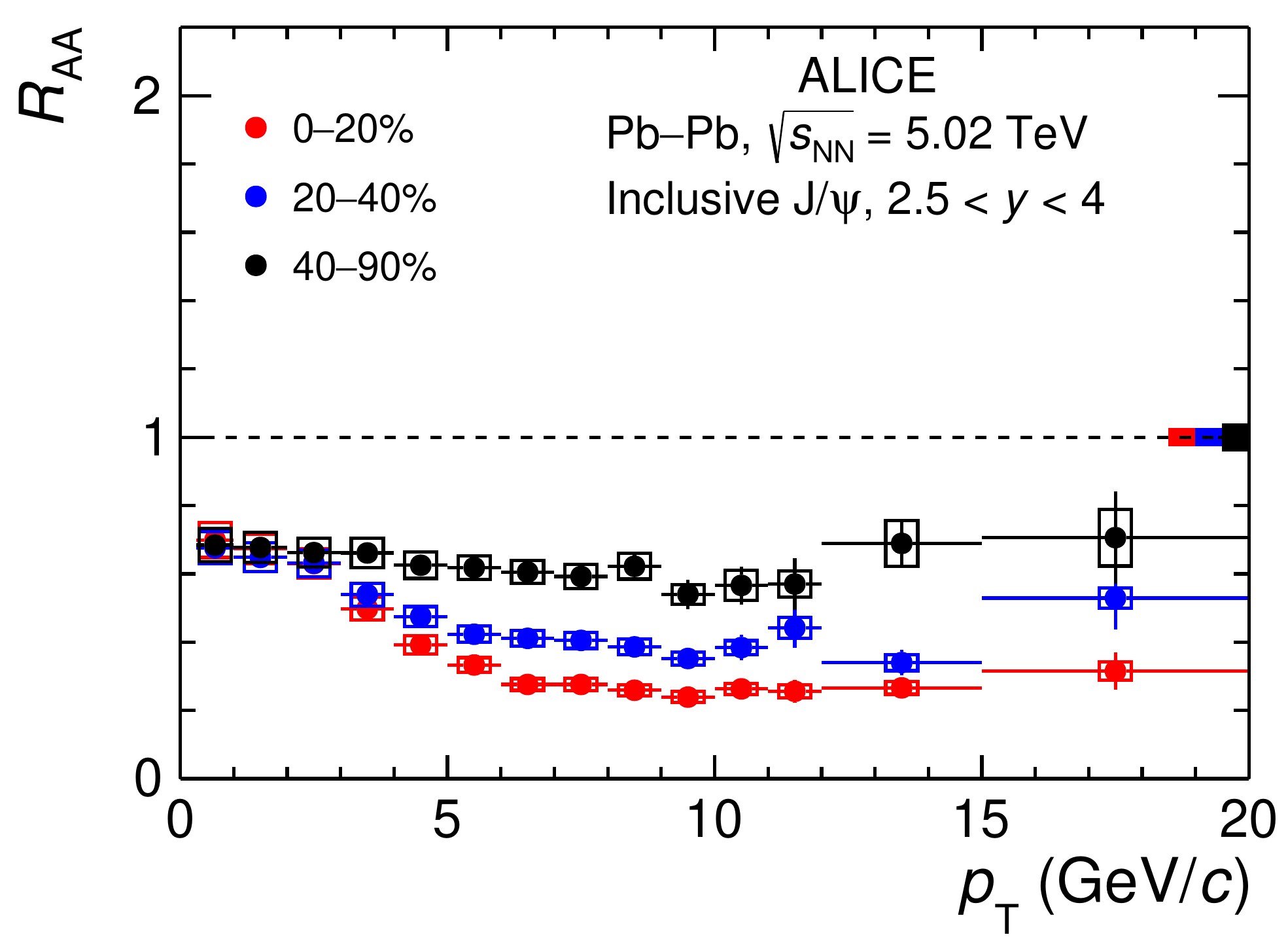}
\caption{Inclusive \jpsi \pt-differential \RAA in Pb--Pb collisions at \fivenn in various centrality intervals. The left panel shows the comparison of \RAA measured in central (0--10\%) and semicentral (30--50\%) collisions at midrapidity. For the data point in the \pt bin 10 $<$ \pt $<$ 15 \GeVc an open symbol is used to highlight the usage of \pp reference from the extrapolation approach. The right panel shows the measured \RAA in three centrality classes, 0--20\%,  20--40\%, and 40--90\%, at forward rapidity. The statistical and systematic uncertainties are indicated, respectively, by the vertical error bars and the open boxes around the data points. The filled boxes around unity show the global uncertainties.}
\label{fig:RAA_vs_pt}
\end{center}
\end{figure}

\begin{figure}[!htb]
\begin{center}
  \includegraphics[width = 7.5 cm]{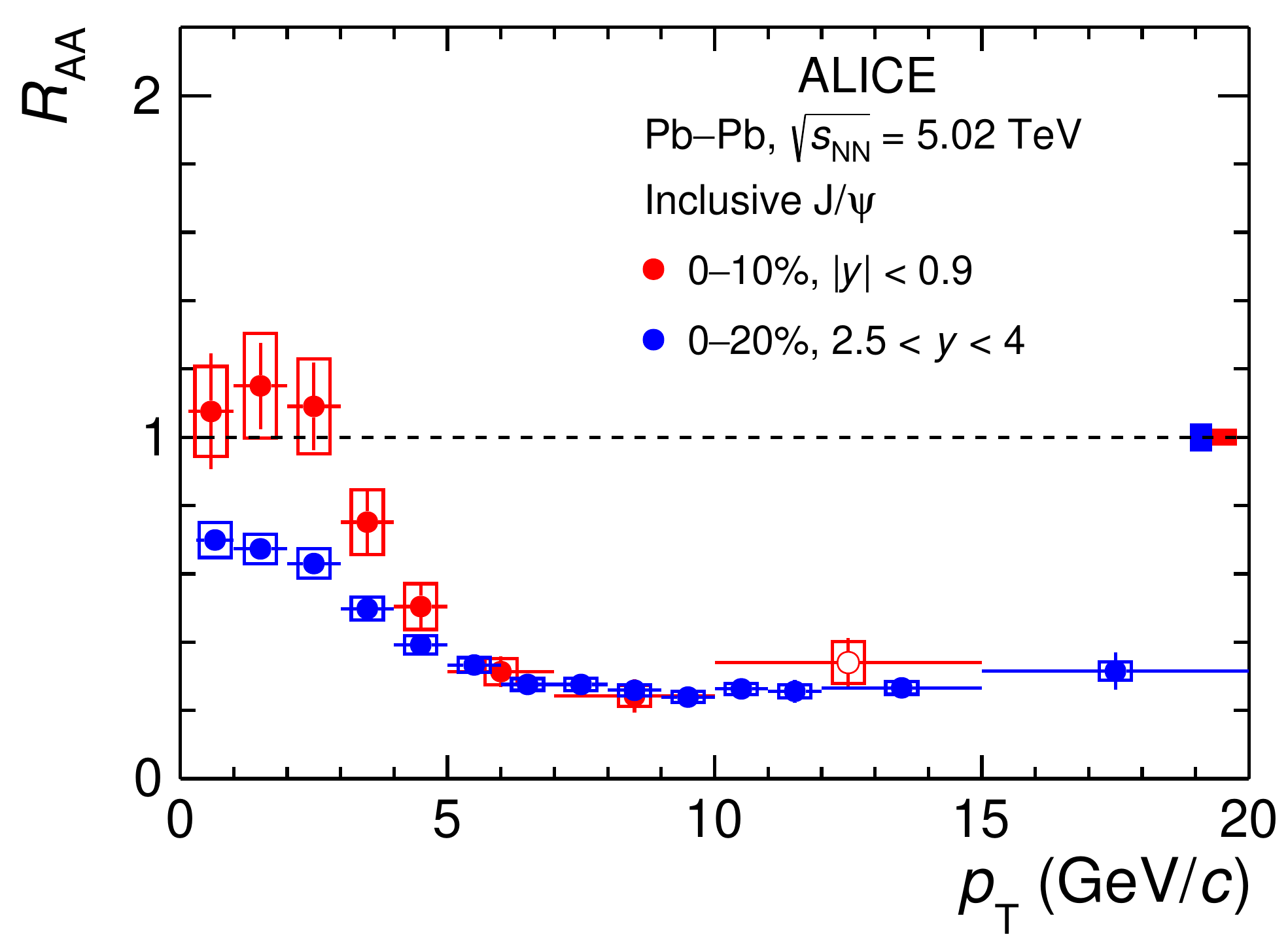}
\caption{Inclusive \jpsi \RAA as a function of \pt in Pb--Pb collisions at \fivenn at midrapidity and forward rapidity, in the 0--10\% centrality class and 0--20\% centrality class, respectively.For the data point in the \pt bin 10 $<$ \pt $<$ 15 \GeVc an open symbol is used to highlight the usage of \pp reference from the extrapolation approach. The statistical and systematic uncertainties are indicated, respectively, by the vertical error bars and the open boxes around the data points. The filled boxes around unity show the global uncertainties.}
\label{fig:RAA_vs_pt_mid_vs_forward_y}
\end{center}
\end{figure}

The \pt-differential \RAA results are shown in Fig.~\ref{fig:RAA_vs_pt} for midrapidity (left panel) and forward rapidity (right panel) in various centrality intervals. The main feature of these measurements is that the \RAA values are relatively large at low \pt ($\pt<5$~\GeVc), in contrast with the strong suppression in the $\pt > 5$~\GeVc range, for central and semicentral collisions. A weaker \pt dependence of the \RAA values is observed from central to peripheral collisions, up to a constant \RAA, within uncertainties, in the 40–90\% centrality interval at forward rapidity. Such a behaviour in the data can be qualitatively understood by the dominance of hot nuclear matter effects for central and semicentral collisions, acting on top of the CNM ones, which were discussed in Refs.~\cite{ALICE:2015sru,ALICE:2021lmn}. In the low-\pt region and in particular in central and semicentral collisions, where the density of charm quarks is larger, the coalescence of charm quark pairs has an important contribution counterbalancing the impact of quarkonium suppression in the QGP. At higher \pt, the \jpsi production is dominated by effects such as dissociation and energy loss, which are expected to be stronger in the most central collisions.
A comparison of the \jpsi \pt-differential \RAA in the most central Pb–Pb collisions at \fivenn between midrapidity and forward rapidity is shown in Fig.~\ref{fig:RAA_vs_pt_mid_vs_forward_y}. Neglecting the slight difference in centrality intervals, the \RAA is higher at midrapidity with respect to forward rapidity at low \pt ($\pt < 3$~\GeVc) with a 2.7$\sigma$ significance, highlighting the strong dependence of \RAA on the local charm quark density, further supporting the picture of quarkonium production via coalescing charm quarks. The two measurements converge to similar values at higher \pt suggesting a weaker dependence on rapidity for the suppression effects.

The measurements presented so far are discussed in the rest of this section in comparison to calculations which employ different approaches in modelling the collision fireball and the hot medium effects having an impact on \jpsi production in \PbPb collisions.

The statistical hadronisation model by Andronic et al.~\cite{Andronic:2019wva} assumes that all charm quarks are produced during the initial hard partonic interactions and then thermalize in the QGP. The relative yields of the charmed hadrons are then determined solely by the equilibrium thermodynamical parameters at the chemical freeze-out, which are fixed from fits to the yields of light-flavoured hadrons. A recent extension of the model~\cite{Andronic:2019wva,Braun-Munzinger:2000csl} implements a hydro-inspired freeze-out hypersurface which allows the calculation of \pt-differential yields in addition to the integrated ones.

The microscopic transport model of Zhuang et al.~\cite{Zhou:2014kka,Chen_2019} implements a real time evolution of the \jpsi, $\psi(2S)$ and $\chi_{\rm c}$ production using a Boltzmann-type rate equation that includes both dissociation and coalescence terms. The dissociation term includes contributions from the temperature dependent colour screening effect and scatterings with thermal partons, i.e. gluon dissociation. The (re)generation of charmonia is implemented by exploiting the detailed balance of the gluon dissociation process. The space--time evolution of the fireball is described using the equations for (2+1)D ideal hydrodynamics. This model includes also the production of non-prompt \jpsi, with the precursor beauty quarks being propagated through the QGP using the Langevin equation.

Similarly to the previously described model, the transport model proposed by Rapp et al. ~\cite{Grandchamp:2001pf,Grandchamp:2003uw,Zhao:2010nk,Zhao:2011cv,Wu:2020zbx}, is also based on a kinetic rate equation to compute the time evolution of charmonium (\jpsi, $\psi(2S)$, and $\chi_c$) yields. The dissociation term of the rate equation employs an inelastic parton scattering cross section of charmonia in the QGP, computed using next-to-leading order perturbative QCD Feynman diagrams, and also includes the effect of in-medium reduced binding energy. The (re)generation rate depends on the charmonium dissociation temperature, which is extracted from lattice QCD calculations, and equilibrium limits computed based on the thermal model. The space--time evolution of heavy-ion collisions is simulated by a cylindrically expanding fireball model with the regenerated charmonium \pt-spectra being calculated in a thermal blast-wave approximation at a temperature and flow velocity reflecting the average production time of each charmonium state~\cite{PhysRevC.48.2462}.

All of the models described above consider the initial state of the nuclear collisions by making assumptions on the total charm quark density produced during the hard partonic collisions and modified by the CNM effects. The two transport models obtain the total charm density based on the measured total charm cross section in pp collisions~\cite{PhysRevD.105.L011103} multiplied by the number of binary nucleon--nucleon collisions. The CNM effects are introduced via different approaches. The microscopic transport model of Rapp et al.~\cite{Grandchamp:2001pf,Grandchamp:2003uw} estimates CNM effects using fits of the measured \pA data, while the transport model of Zhuang et al.~\cite{Zhao:2010nk,Zhao:2011cv} uses the nuclear parton distribution functions and their uncertainties from EPS09~\cite{Eskola:2009uj}.
The SHMc extracts the total charm cross section from the ALICE measurements of D meson production in \PbPb collisions~\cite{alicedmesons}. The large uncertainty on the estimation of CNM effects is inherited by these model calculations.

 At large \pt, the fragmentation of high-energy partons may become the main mechanism for \jpsi production. In that case, energy loss of partons due to multiple scattering in the QGP leads to \jpsi suppression at high \pt. In the model by Arleo et al.~\cite{Arleo:2017ntr}
, the quenching of large-\pt particles ($\pt > 10$~\GeVc) is assumed to be mostly due to radiative parton energy loss. In this approach, the \pt dependence of \RAA is fully predicted from the model proposed by Baier et al.~\cite{Baier:1996kr,Baier:1996sk}, which employed a medium-induced gluon spectrum. The \RAA value is computed from the mean energy loss, which is extracted from a fit to the charged hadron \RAA, measured in various collision systems, the average fragmentation function, and the colour coupling factor of the parton. At forward rapidity, the mean energy loss is further corrected for the charged-hadron multiplicity difference between midrapidity and forward rapidity. The model uncertainties arise from the uncertainties on these inputs. This model does not include the production of non-prompt \jpsi, but the \RAA variation, when accounting for this contribution, is expected to lie within the theoretical uncertainties.  

\begin{figure}[!htb]
\begin{center}
  \includegraphics[width = 7.5 cm]{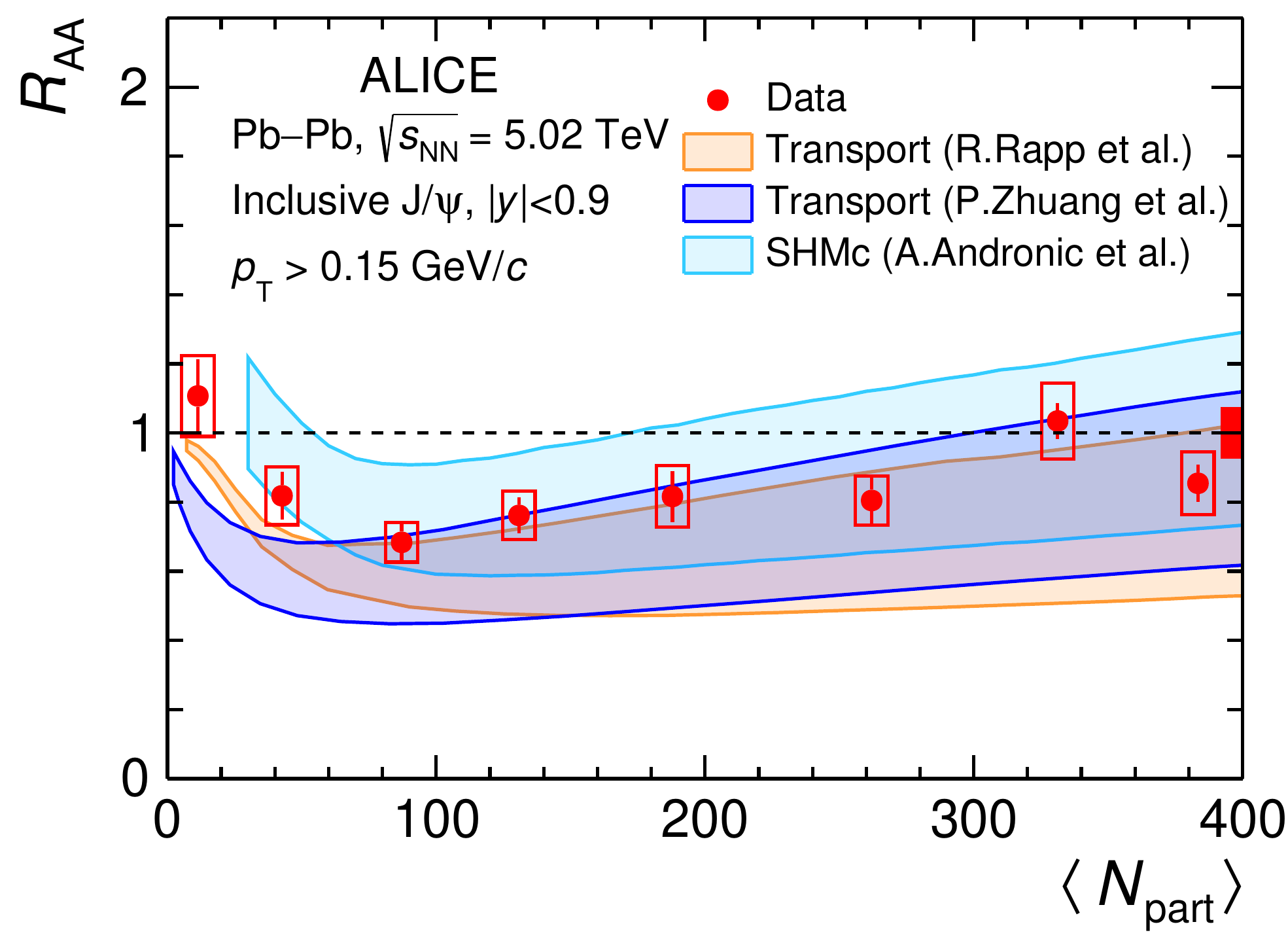}
\caption{Inclusive \jpsi \RAA at midrapidity, integrated over \pt, as a function of \Npart in Pb--Pb collisions at \fivenn and compared to model calculations from Refs.~\cite{Andronic:2019wva,Zhao:2007hh,Zhou:2014kka}. 
}
\label{fig:Raa_vs_cent_models}
\end{center}
\end{figure}

The \pt-integrated nuclear modification factor measured in \PbPb collisions at \fivenn at midrapidity is shown in Fig.~\ref{fig:Raa_vs_cent_models} in comparison with results from the SHMc and the two transport-model calculations. The calculations are shown as coloured bands, illustrating the uncertainties on the initial effects, mainly CNM effects, described above. Within the model uncertainties, all three predictions agree with the data. One can note though that the data lie on the upper edge of the transport-model calculations, while they are in good agreement with the central values from the SHMc calculations for semicentral and central collisions.

\begin{figure}[!htb]
\begin{center}
  \includegraphics[width = 7.5 cm]{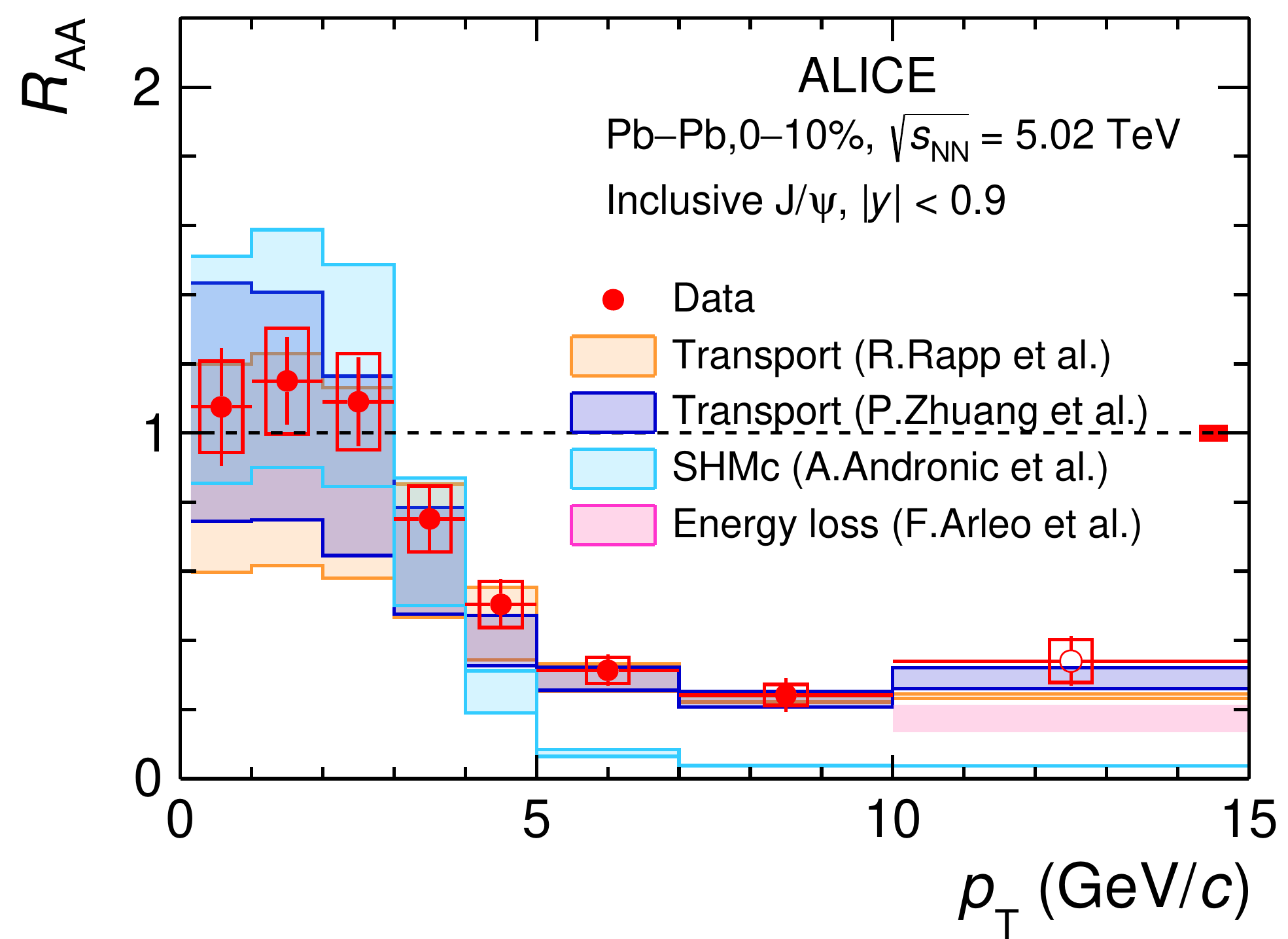}
  \includegraphics[width = 7.5 cm]{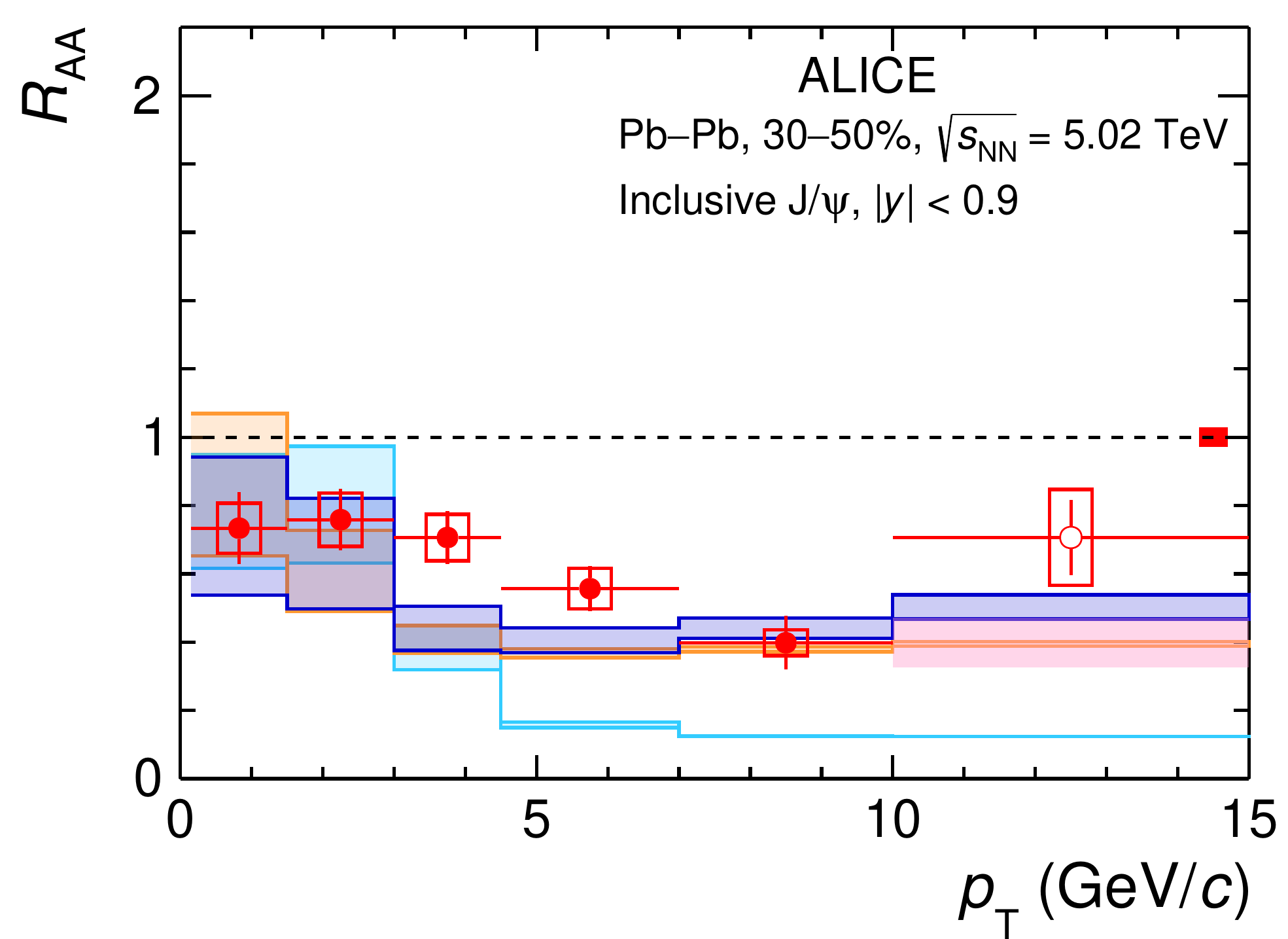}
\caption{Transverse-momentum dependence of the \jpsi \RAA in $\PbPb$ collisions at \fivenn
at midrapidity in the $0$--$10$\% (left panel) and $30$--$50$\% (right panel) centrality intervals. For the data point in the \pt bin 10 $<$ \pt $<$ 15 \GeVc an open symbol is used to highlight the usage of \pp reference from the extrapolation approach. The data are compared with model calculations from Refs.~\cite{Andronic:2019wva,Zhao:2007hh,Zhou:2014kka,Arleo:2017ntr}.}
\label{fig:RAA_vs_pt_model}
\end{center}
\end{figure}

\begin{figure}[!htb]
\begin{center}
  \includegraphics[width = 7.5 cm]{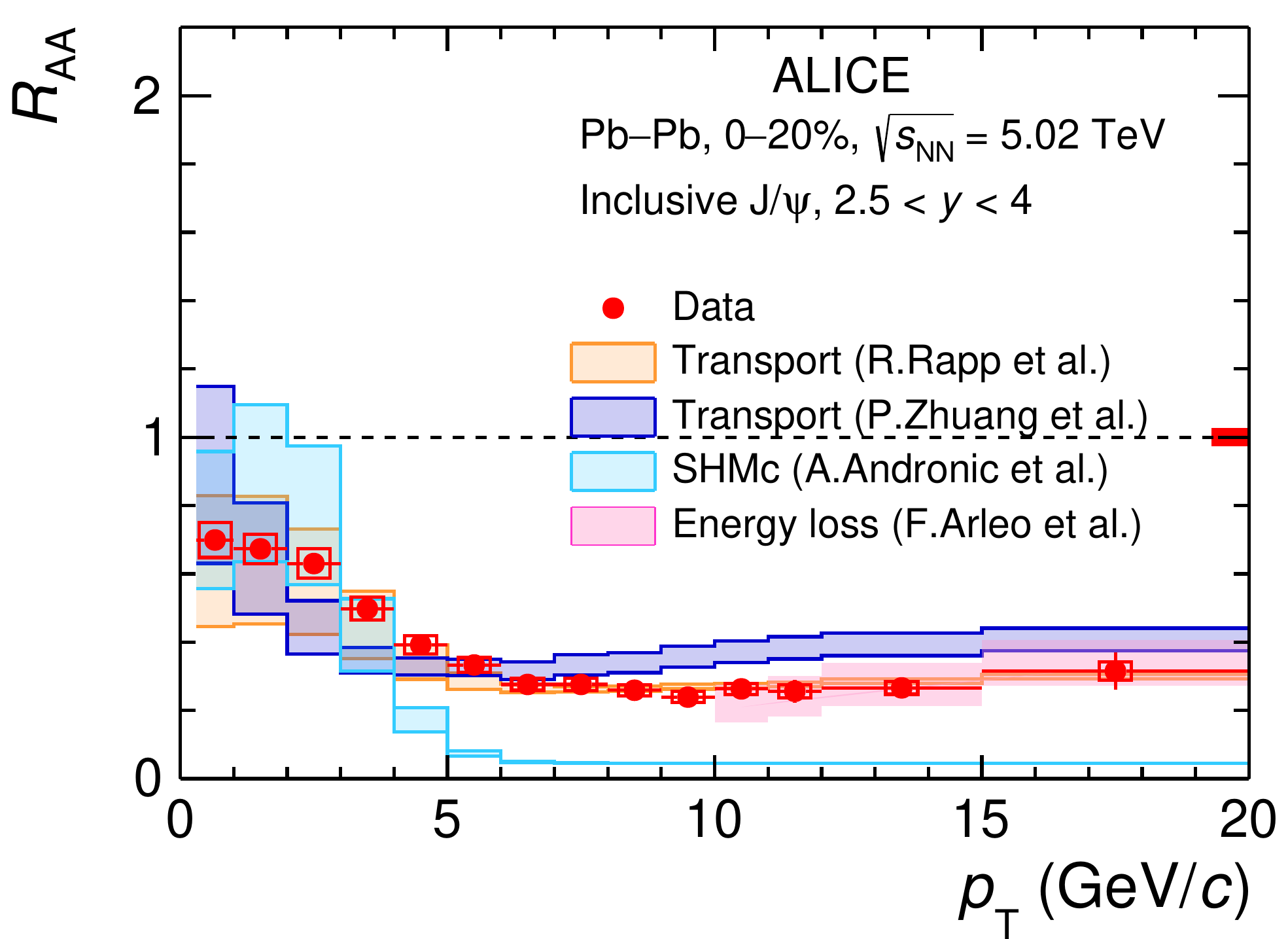}
  \includegraphics[width = 7.5 cm]{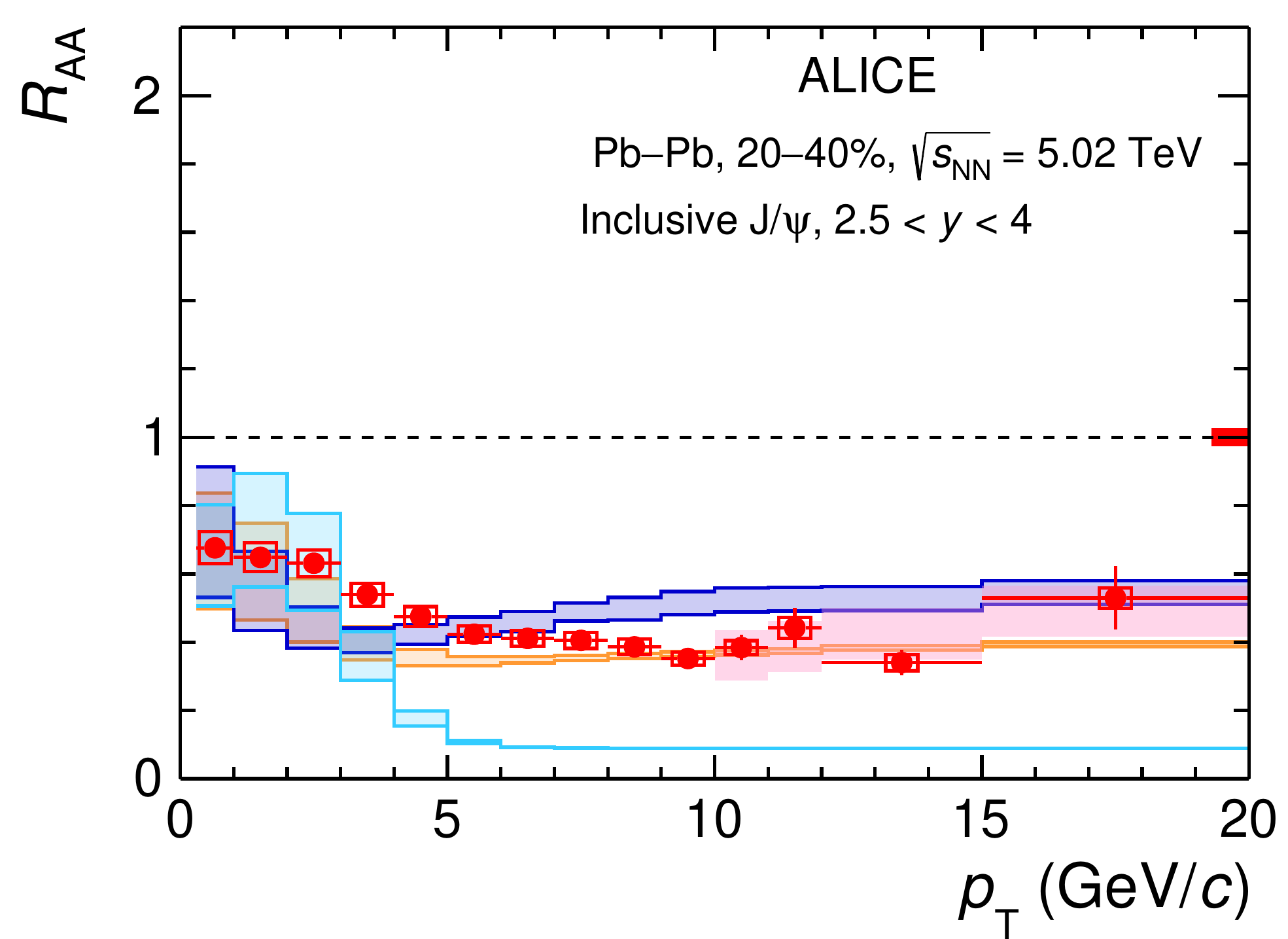}
   \includegraphics[width = 7.5 cm]{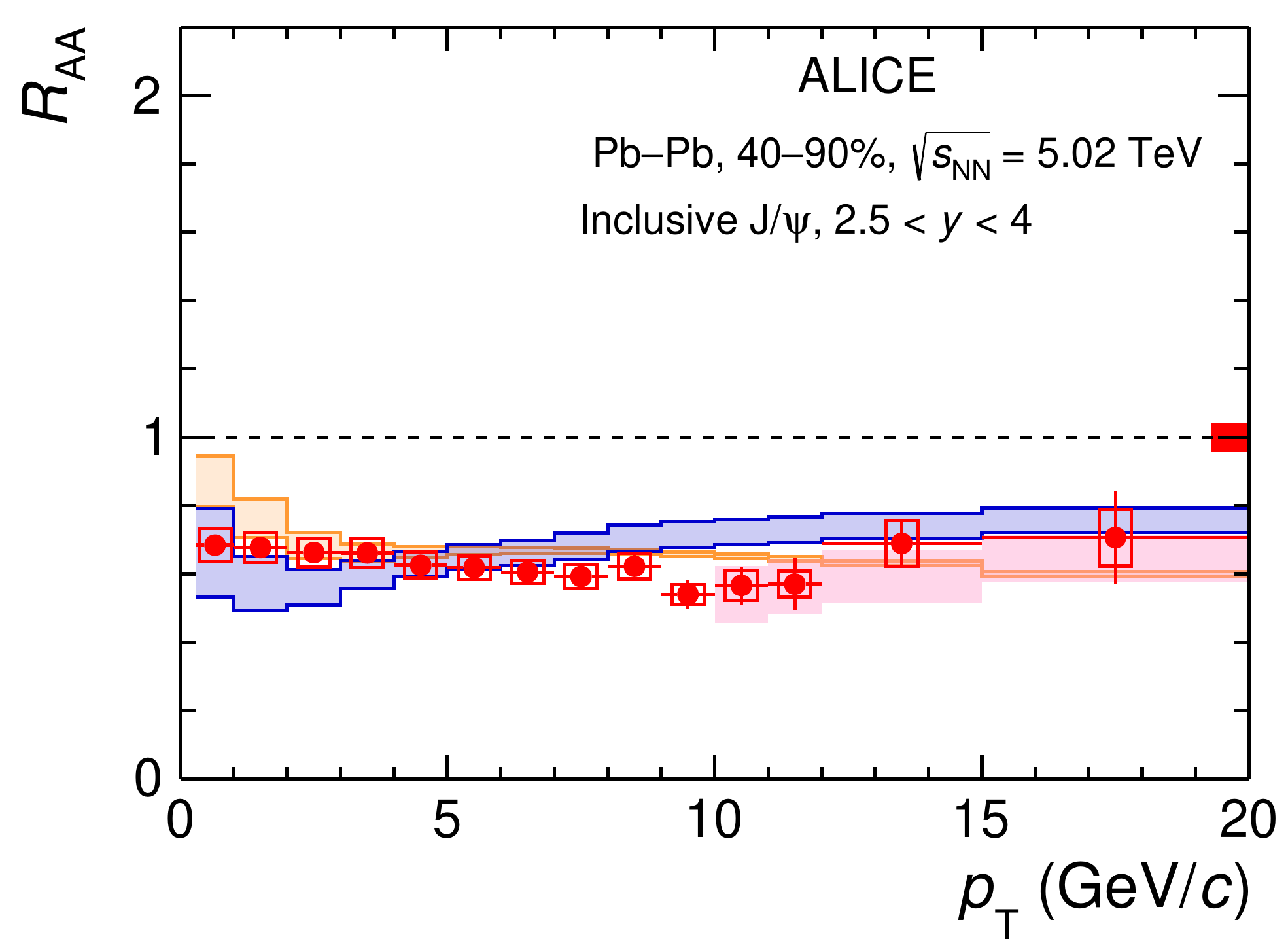}
\caption{Transverse-momentum dependence of the \jpsi \RAA at forward rapidity in the $0$--$20$\%, $20$--$40$\% and $40$--$90$\%  centrality intervals. The data are compared with model calculations from Refs.~\cite{Andronic:2019wva,Zhao:2007hh,Zhou:2014kka,Arleo:2017ntr}}
\label{fig:RAA_vs_pt_model_forward}
\end{center}
\end{figure}

Figures~\ref{fig:RAA_vs_pt_model} and ~\ref{fig:RAA_vs_pt_model_forward} show the \pt-differential \RAA measurements for various centrality intervals at midrapidity and forward rapidity, respectively, in comparison with the available model calculations. With the exception of the energy-loss calculations, available only for $\pt>10$~\GeVc, all of the models cover the full \pt range in which these measurements were performed. The SHMc model calculations are in good agreement with the data at low \pt at both midrapidity and forward rapidity. However, the \RAA is underestimated for $\pt>5$~\GeVc in all centrality intervals in both rapidity ranges. 
This might be attributed to physical sources missing in this approach, such as the contributions from surviving primordial \jpsi or non-prompt \jpsi from beauty-hadron decays, but also to an underestimated amount of radial flow acquired by the charm quarks during the system evolution. 
 A similar conclusion can also be drawn from the comparison of the \pt-differential yields in \PbPb collisions shown in Fig.~\ref{fig:JpsiPtSpectra_mid} where the measured spectrum is harder than the one from the SHMc calculations. The two transport models are in better quantitative agreement with data than the SHMc model. Both of the transport models provide a good description of the \RAA at both low and high \pt. However, the model calculations in the low-\pt region, where \jpsi production is dominated by coalescence in these models, do not describe the detailed shape of the \pt dependence of \RAA, in particular in semicentral collisions, which points to a still not perfectly understood dynamics of charm-quark coalescence.

The energy-loss calculations by Arleo et al~\cite{Arleo:2017ntr}, performed in all studied centrality ranges for $\pt>10$~\GeVc, are in good agreement with the measurements, which, based on the model assumptions, suggests that the dominant mechanism in this kinematic regime is indeed energy loss, similar to that of the other hadrons measured at LHC energies.

\FloatBarrier

\subsection{The inclusive \textbf{J/$\psi$} \meanpt and \meanptsq } \label{subsec:meanpt}

\begin{figure}[!htb]
\begin{center}
  \includegraphics[width = 7.5 cm]{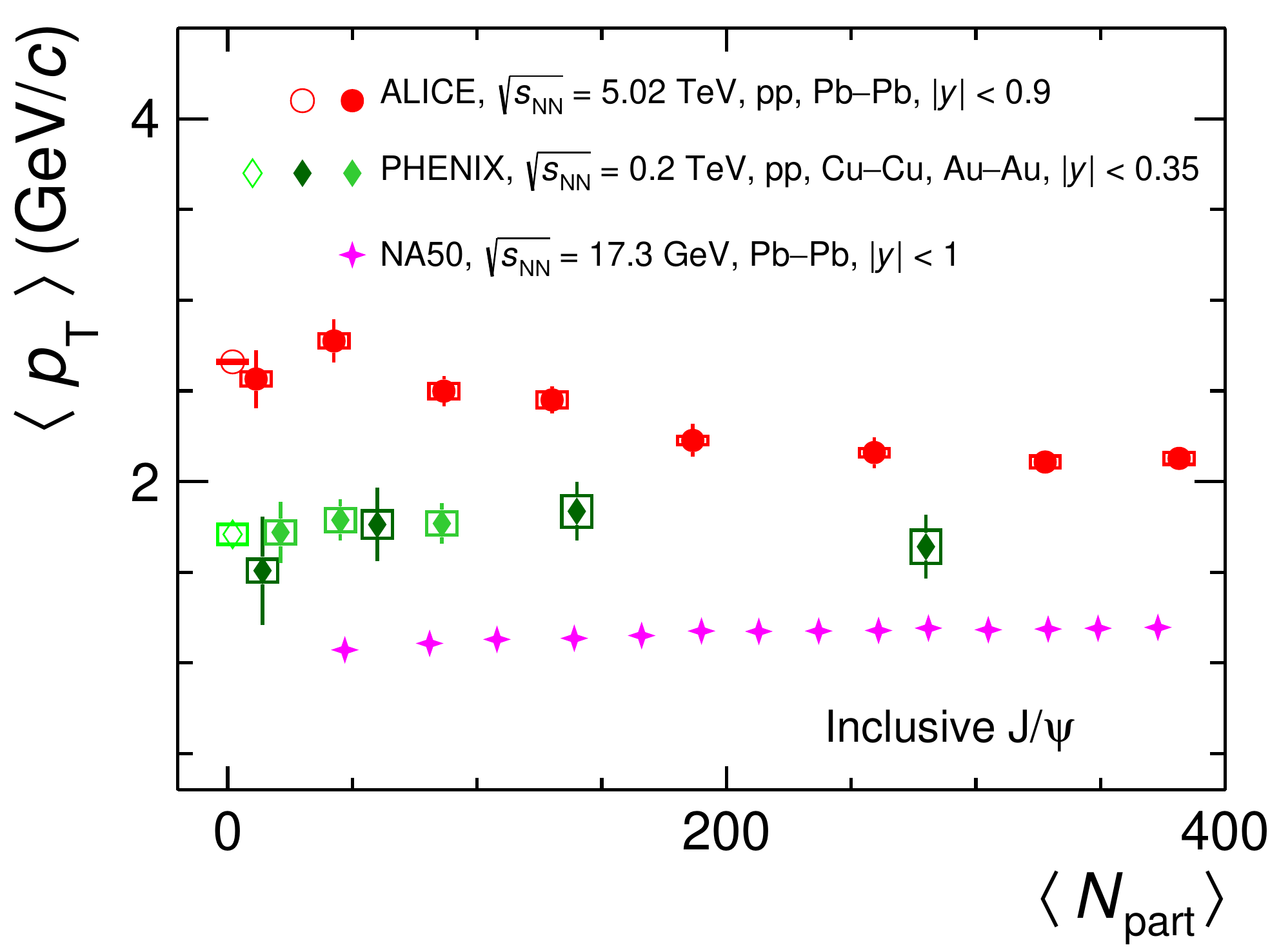}
  \includegraphics[width = 7.5 cm]{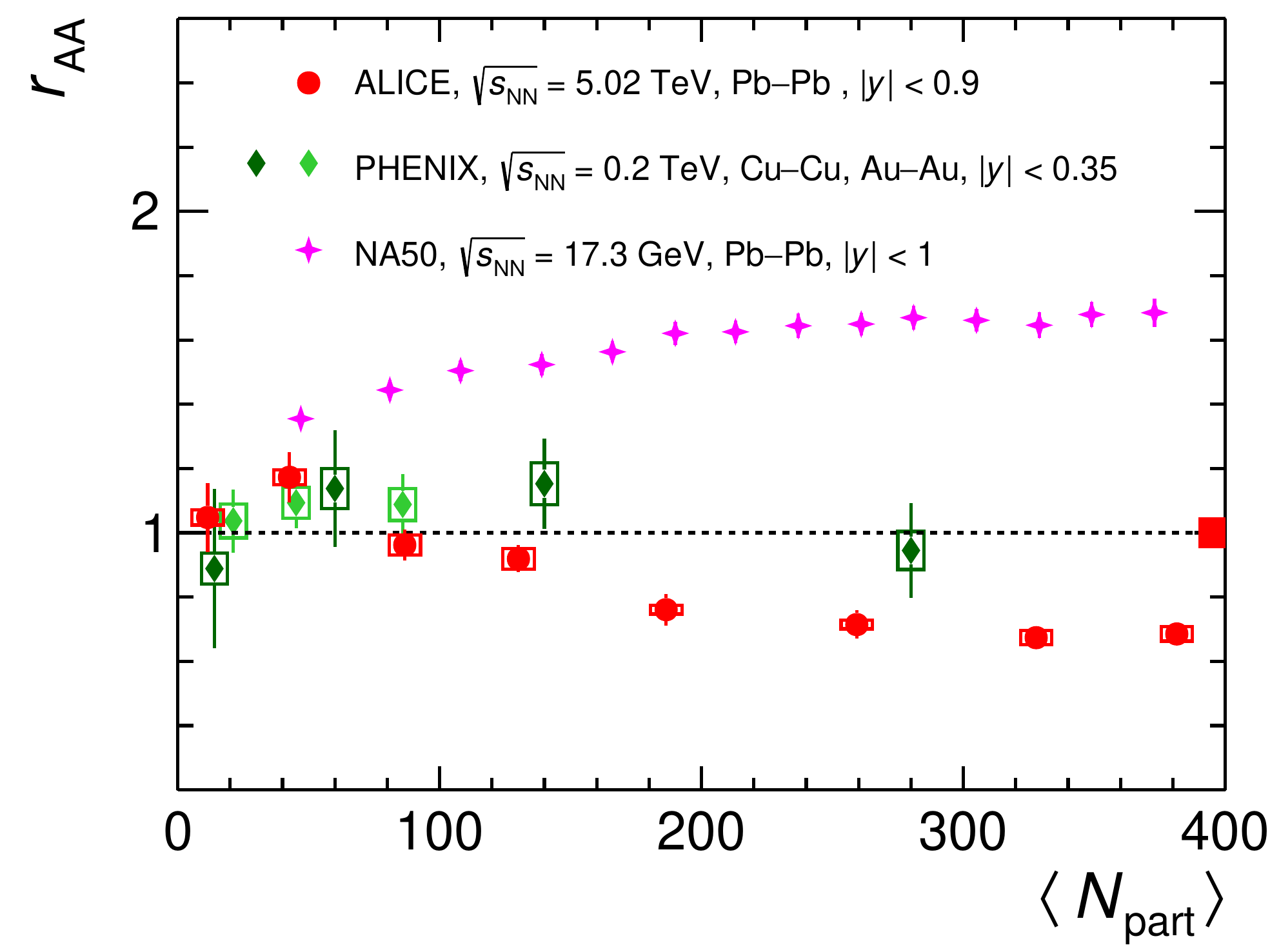}
\caption{Left panel: Inclusive \jpsi \meanpt as a function of the mean number of participants in Pb--Pb collisions at \fivenn at midrapidity. Right panel: Inclusive \jpsi \rAA as a function of centrality at \fivenn and compared with measurements at lower energies from RHIC~\cite{PHENIX:2006gsi,Adare:2008sh,Adare:2011vq} and the SPS~\cite{Abreu:2000xe}. The statistical and systematic uncertainties are indicated, respectively, by the vertical error bars and the open boxes around the data points. The filled box around unity on the right panel shows the global uncertainty.}
\label{fig:mean_pt_rAA_with_data}
\end{center}
\end{figure}

Observables that allow for a more differential study of the \jpsi \pt spectrum with respect to the centrality of the collisions are the \jpsi \meanpt and \meanptsq. The latter is typically quantified using the \rAA, defined in Eq.\ref{eq:little_raa}, which is related to the broadening or narrowing of the \jpsi \pt spectrum relative to that in \pp collisions. The \meanpt and the \rAA measured at midrapidity are shown in Fig.~\ref{fig:mean_pt_rAA_with_data}  in the left and right panels, respectively, as a function of $\langle\Npart\rangle$. Similar measurements were done at the forward rapidity as well~\cite{ALICE:2019lga}. These results are compared with similar measurements in heavy-ion collisions at RHIC~\cite{PHENIX:2006gsi,PhysRevLett.101.122301,Adare:2011vq} and SPS~\cite{Abreu:2000xe} energies. The \meanpt results are also compared with the value reported by the ALICE Collaboration in \pp collisions at midrapidity at $\s=5.02$~TeV~\cite{Acharya:2019lkw}, showing a good agreement with the value measured in the most peripheral \PbPb collisions. The data show that for a given \Npart, the \jpsi \meanpt grows with increasing collision energy. However, while the data indicate no centrality dependence of the \meanpt at the SPS and RHIC energies, a monotonically decreasing trend from the most peripheral to the most central collisions is observed in the ALICE measurements, which reflects the gradual increase of the low \pt (re)generation component. The \rAA results support the observations for the \meanpt, showing a decrease from unity towards central collisions for the ALICE measurements. The RHIC results are compatible with unity over the whole covered centrality range, while the SPS data indicate a strong increase from peripheral to central collisions, suggesting that CNM effects such as the Cronin effect ~\cite{PhysRevLett.88.232303} have an impact on the \jpsi \pt shape.

\begin{figure}[!htb]
\begin{center}
  \includegraphics[width = 7.5 cm]{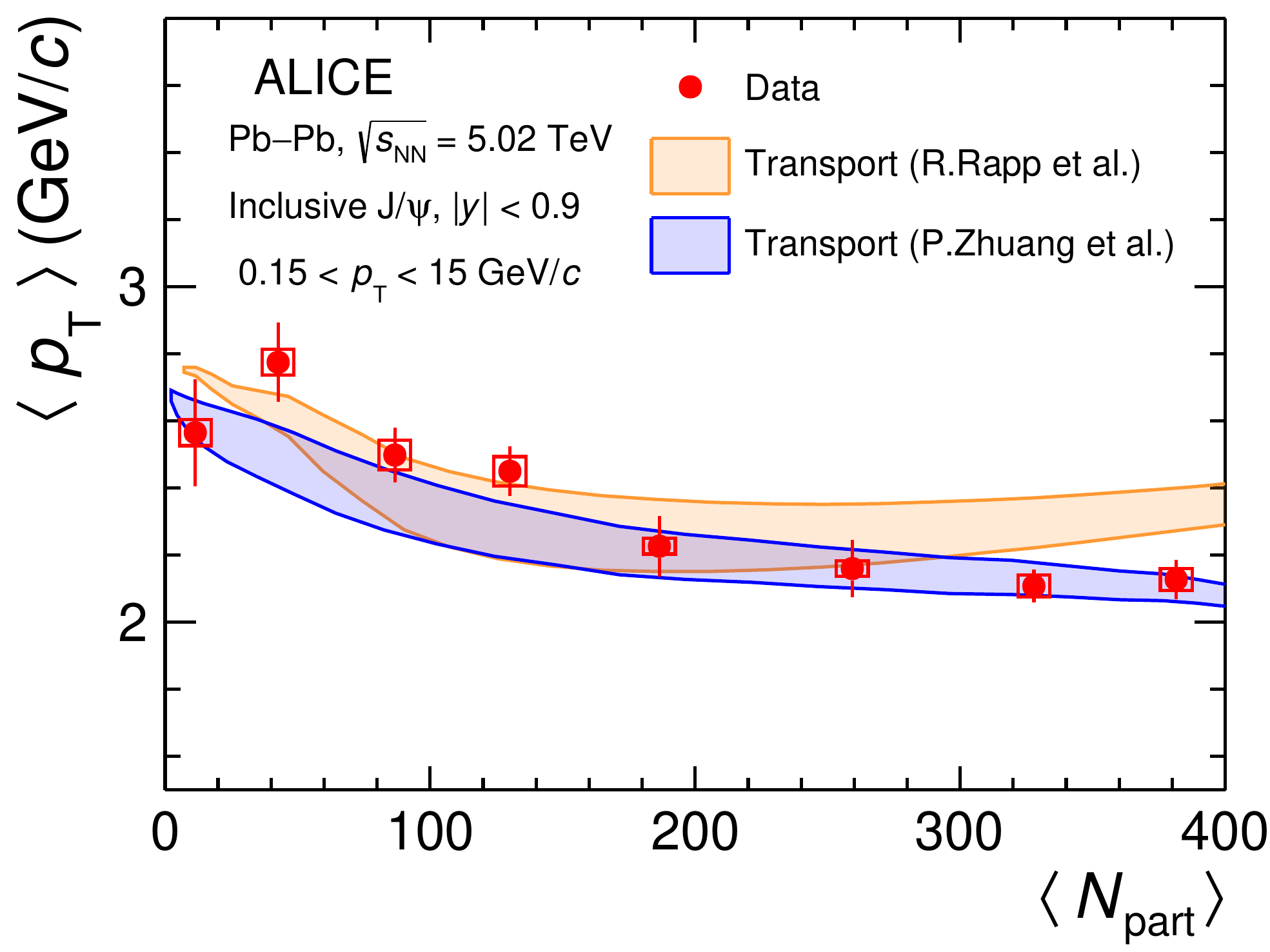}
  \includegraphics[width = 7.5 cm]{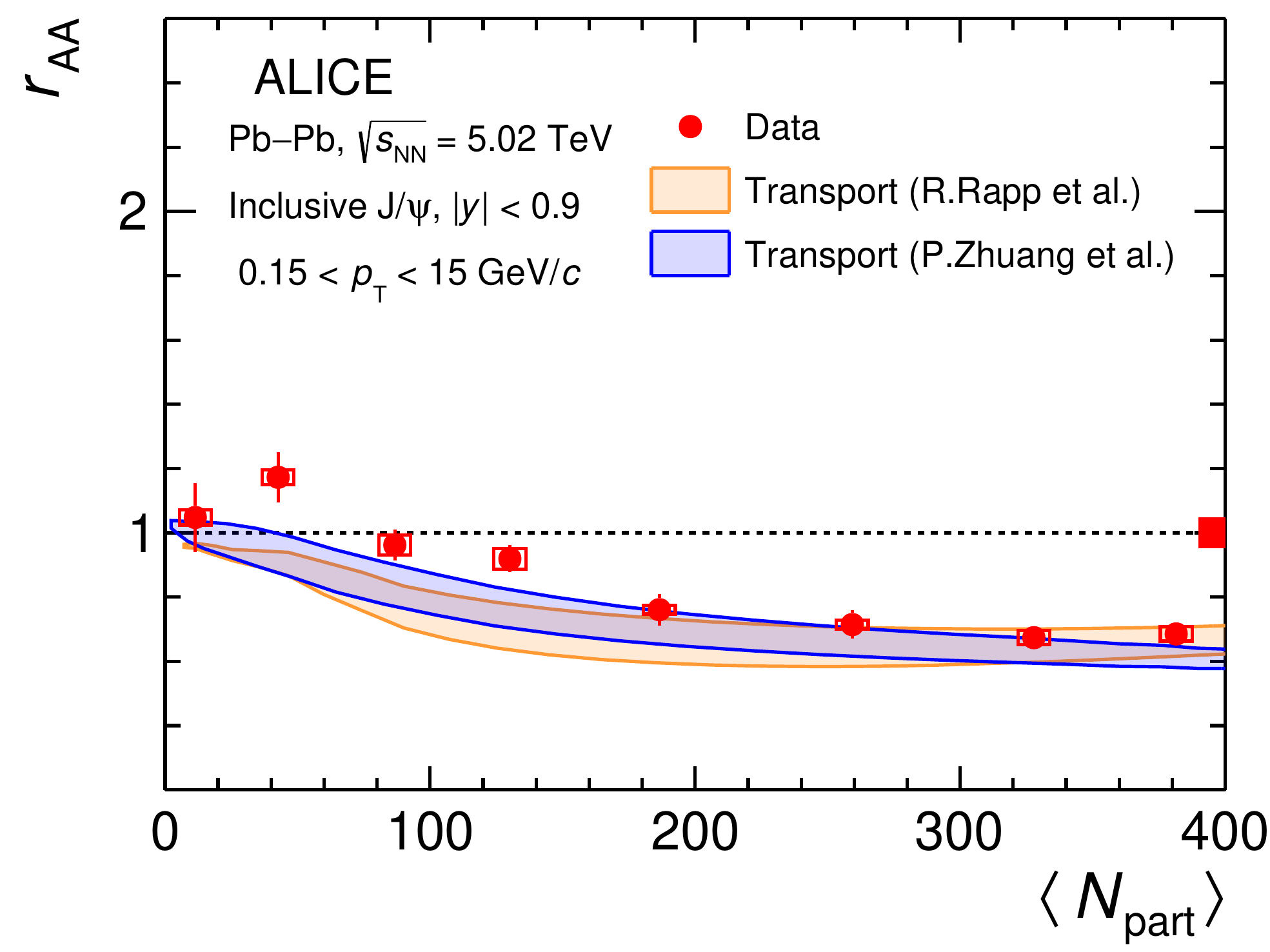}
\caption{Inclusive \jpsi \meanpt as a function of the mean number of participants in Pb--Pb collisions at \fivenn at midrapidity (left panel), and \jpsi \rAA as a function of centrality (right panel).
The results are compared with transport model calculations~\cite{Zhao:2007hh,Zhou:2014kka}}
\label{fig:mean_pt_rAA_with_model}
\end{center}
\end{figure}

The \meanpt and \meanptsq results are also compared with the aforementioned transport model calculations, which show a good agreement with the trends observed in data as demonstrated in Fig.~\ref{fig:mean_pt_rAA_with_model}. Although overall in good quantitative agreement, the calculations by Rapp et al. overestimate the \jpsi \meanpt for the most central collisions, while the \meanptsq results are slightly underestimated by both models in the semicentral and peripheral range ($50<\Npart<150$).

\FloatBarrier

\subsection{The \textbf{J/$\psi$} to D$^{0}$ yield ratio} 
\label{subsec:jpsi_D0_ratio}

A long awaited measurement which helps understanding the details of the \jpsi production in heavy-ion collisions is the ratio between the \jpsi and the \dzero yields, both measured in the same collision system. Such a measurement provides a tight constraint to models because some of the model parameters and most model uncertainties related to the \ccbar cross section
cancel in the ratio. This ratio is sensitive to the hadronisation mechanisms of the different charm hadrons. While a model independent measurement of the \ccbar production density in heavy-ion collisions is not available, it is still useful to compare the \jpsi yield with the recently published ALICE measurements of the \dzero yield down to zero \pt~\cite{alicedmesons}.

\begin{figure}[!htb]
\begin{center}
  \includegraphics[width = 8.5 cm]{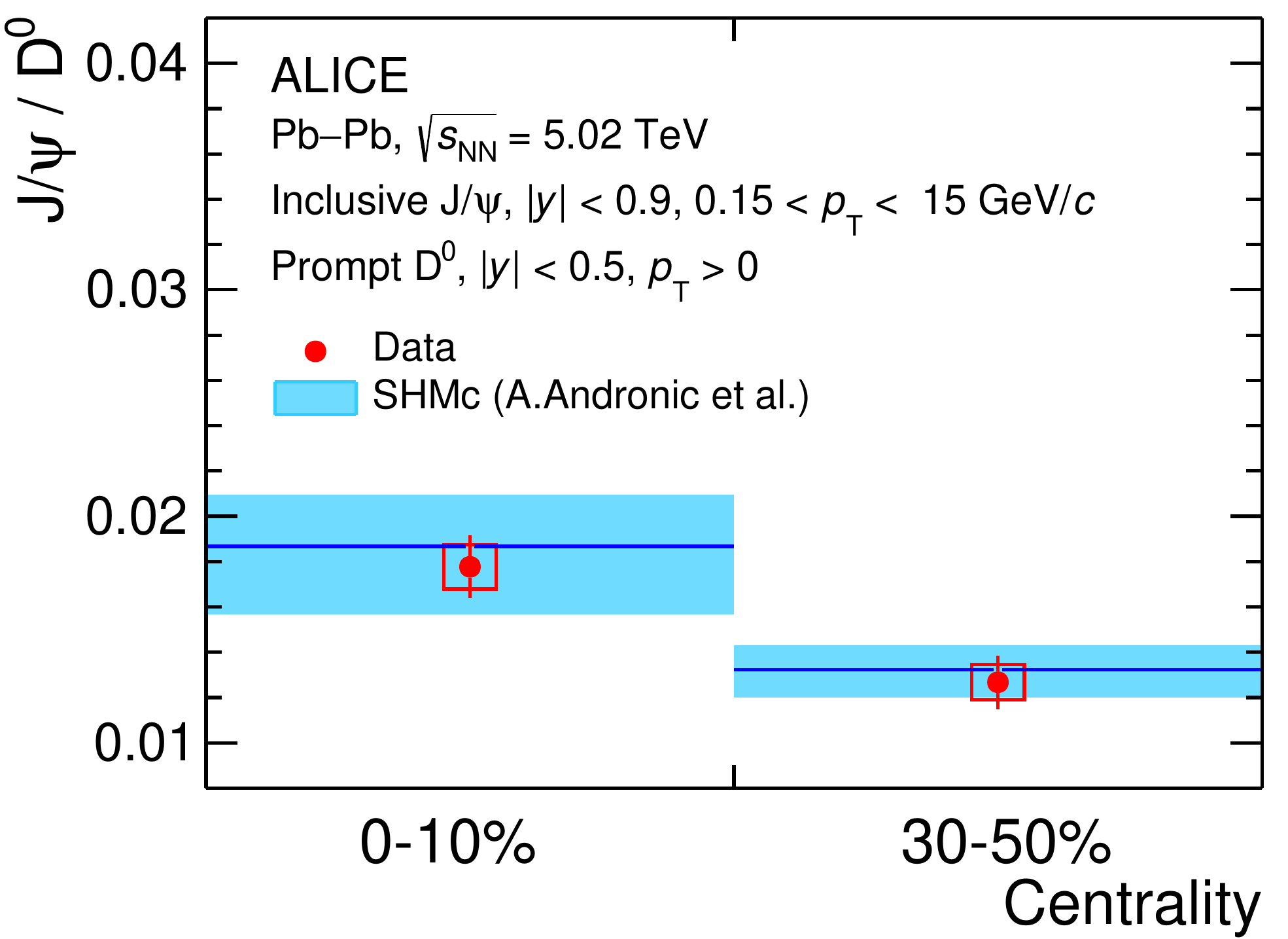}
\caption{Inclusive \jpsi to $D^{0}$ yield~\cite{alicedmesons} ratio at \fivenn at midrapidity for the 0--10\% and 30--50\% centrality intervals. Vertical lines and open boxes represent the statistical and systematical uncertainties, respectively. The measurements are compared with SHMc model predictions~\cite{Andronic:2019wva}. }
\label{fig:D0tojpsi}
\end{center}
\end{figure}

Figure~\ref{fig:D0tojpsi} shows the measured \pt-integrated \jpsi to \dzero yield ratio in central (0--10\%) and semicentral (30--50\%) collisions. The largest source of systematic uncertainty for both measurements comes from tracking efficiency and it is considered correlated between the \dzero and \jpsi measurements, and consequently cancels in the ratio. Without this uncertainty, the numerical values of the \jpsi yields are $0.12 \pm 0.005$~(stat.)~$\pm 0.012$~(syst.)~and $0.016 \pm 0.0008$~(stat.)~$\pm 0.0004$~(syst.) for (0--10\%) and (30--50\%) centrality intervals, respectively. 
The statistical (systematical) uncertainty of the ratio is the quadratic sum of the statistical (systematical) uncertainties of the two measurements. 
The results suggest a higher value for this ratio in central compared to semicentral collisions. This is supported by the SHMc calculations~\cite{Andronic:2019wva}, which suggests both the \jpsi and \dzero are produced via the coalescence of charm quarks at the phase boundary, the ratio being determined by the charm fugacity. The SHMc model gives a good description of the data. The model uncertainty from the SHMc model is due to uncertainties on the charm fugacity parameter, which is fitted to the ALICE D$^{0}$ data ~\cite{alicedmesons}.
 
\FloatBarrier

\section{Conclusions}
\label{Conclusions} 
The \jpsi \pt-differential yields and nuclear modification factors \RAA 
measured from $\pt = 0.15$~GeV/c up to $15$~GeV/c in the 0–10\% and 30–50\% centrality ranges at midrapidity and from $\pt = 0.3$~GeV/c up to $\pt = 20$~GeV/c in the 0--20\%, 20--40\% and 40--90\% centrality ranges at forward rapidity, and the centrality dependent \jpsi \meanpt and \rAA  measured at midrapidity, are reported and discussed in comparison with model calculations.

The centrality dependent \pt-integrated \RAA in peripheral collisions shows similar values at both midrapidity and forward rapidity, while a hint of an increasing trend of the \RAA is observed at midrapidity towards central collisions. When looking at the \pt-differential \RAA, relatively large values are observed at low \pt, which are compatible with unity for $\pt<3$~\GeVc in the 0--10\% centrality interval at $|y|<0.9$, while a strong nuclear suppression is seen at higher \pt in central and semicentral collisions. In addition, \RAA is higher at midrapidity 
than at forward rapidity  for $\pt < 3$~GeV/c in the most central collisions. A weaker \pt dependence of the \RAA values is observed for more peripheral collisions. Such a behaviour, for the central and semicentral collisions, can be explained by a large contribution from (re)generation to the \jpsi yields. This is supported by the statistical hadronisation model and by two microscopic transport model calculations when compared to the data. However, the large model uncertainties, which are mainly due to the assumptions on the collision initial conditions, prevent from drawing a clear conclusion on the phenomenology of the \jpsi production in heavy-ion collisions at LHC energies.
The \jpsi nuclear modification factor at high \pt is well described by the transport models and also by a \jpsi energy loss model, while it is largely underestimated in the hydro-inspired freeze-out approach implemented together with the SHMc model. For most central events, the \RAA converge to similar values at high \pt at mid and forward rapidity suggesting a weaker dependence on rapidity for the suppression effects. The centrality dependent \jpsi \meanpt and \rAA measurements are compared with similar results at lower energies from RHIC~\cite{PHENIX:2006gsi,Adare:2008sh,Adare:2011vq} and SPS~\cite{Abreu:2000xe}, with the centrality trends showing an opposite behaviour between the LHC and the lower energy results. This behaviour is compatible with a strong contribution from the regeneration component which tends to soften the \pt distributions. The two microscopic transport models describe the data within the uncertainties.

The ratio of inclusive \pt-integrated \jpsi to \dzero yields measured by ALICE at midrapidity in \PbPb collisions is presented for the first time in the 0--10\% and 30--50\% centrality ranges. The data shows a larger value of this ratio in 0--10\% compared to 30--50\% collisions, which is in good agreement with the expectation from the SHMc model that fixes the charm fugacity parameter based on the \dzero yields measured by ALICE. 

The large improvements in experimental accuracy expected for the LHC Run 3 and 4 for charmonium measurements and more in general for heavy-quark production, the improved measurements of total charm quark cross section and CNM effects are critical to constrain the phenomenological model calculations, will allow to settle the longstanding questions regarding the mechanisms behind charmonium production in heavy-ion collisions.


\newenvironment{acknowledgement}{\relax}{\relax}
\begin{acknowledgement}
\section*{Acknowledgements}
We would like to thank the theory groups who kindly provided us with their calculations and for the fruitful discussions.


The ALICE Collaboration would like to thank all its engineers and technicians for their invaluable contributions to the construction of the experiment and the CERN accelerator teams for the outstanding performance of the LHC complex.
The ALICE Collaboration gratefully acknowledges the resources and support provided by all Grid centres and the Worldwide LHC Computing Grid (WLCG) collaboration.
The ALICE Collaboration acknowledges the following funding agencies for their support in building and running the ALICE detector:
A. I. Alikhanyan National Science Laboratory (Yerevan Physics Institute) Foundation (ANSL), State Committee of Science and World Federation of Scientists (WFS), Armenia;
Austrian Academy of Sciences, Austrian Science Fund (FWF): [M 2467-N36] and Nationalstiftung f\"{u}r Forschung, Technologie und Entwicklung, Austria;
Ministry of Communications and High Technologies, National Nuclear Research Center, Azerbaijan;
Conselho Nacional de Desenvolvimento Cient\'{\i}fico e Tecnol\'{o}gico (CNPq), Financiadora de Estudos e Projetos (Finep), Funda\c{c}\~{a}o de Amparo \`{a} Pesquisa do Estado de S\~{a}o Paulo (FAPESP) and Universidade Federal do Rio Grande do Sul (UFRGS), Brazil;
Bulgarian Ministry of Education and Science, within the National Roadmap for Research Infrastructures 2020-2027 (object CERN), Bulgaria;
Ministry of Education of China (MOEC) , Ministry of Science \& Technology of China (MSTC) and National Natural Science Foundation of China (NSFC), China;
Ministry of Science and Education and Croatian Science Foundation, Croatia;
Centro de Aplicaciones Tecnol\'{o}gicas y Desarrollo Nuclear (CEADEN), Cubaenerg\'{\i}a, Cuba;
Ministry of Education, Youth and Sports of the Czech Republic, Czech Republic;
The Danish Council for Independent Research | Natural Sciences, the VILLUM FONDEN and Danish National Research Foundation (DNRF), Denmark;
Helsinki Institute of Physics (HIP), Finland;
Commissariat \`{a} l'Energie Atomique (CEA) and Institut National de Physique Nucl\'{e}aire et de Physique des Particules (IN2P3) and Centre National de la Recherche Scientifique (CNRS), France;
Bundesministerium f\"{u}r Bildung und Forschung (BMBF) and GSI Helmholtzzentrum f\"{u}r Schwerionenforschung GmbH, Germany;
General Secretariat for Research and Technology, Ministry of Education, Research and Religions, Greece;
National Research, Development and Innovation Office, Hungary;
Department of Atomic Energy Government of India (DAE), Department of Science and Technology, Government of India (DST), University Grants Commission, Government of India (UGC) and Council of Scientific and Industrial Research (CSIR), India;
National Research and Innovation Agency - BRIN, Indonesia;
Istituto Nazionale di Fisica Nucleare (INFN), Italy;
Japanese Ministry of Education, Culture, Sports, Science and Technology (MEXT) and Japan Society for the Promotion of Science (JSPS) KAKENHI, Japan;
Consejo Nacional de Ciencia (CONACYT) y Tecnolog\'{i}a, through Fondo de Cooperaci\'{o}n Internacional en Ciencia y Tecnolog\'{i}a (FONCICYT) and Direcci\'{o}n General de Asuntos del Personal Academico (DGAPA), Mexico;
Nederlandse Organisatie voor Wetenschappelijk Onderzoek (NWO), Netherlands;
The Research Council of Norway, Norway;
Commission on Science and Technology for Sustainable Development in the South (COMSATS), Pakistan;
Pontificia Universidad Cat\'{o}lica del Per\'{u}, Peru;
Ministry of Education and Science, National Science Centre and WUT ID-UB, Poland;
Korea Institute of Science and Technology Information and National Research Foundation of Korea (NRF), Republic of Korea;
Ministry of Education and Scientific Research, Institute of Atomic Physics, Ministry of Research and Innovation and Institute of Atomic Physics and Universitatea Nationala de Stiinta si Tehnologie Politehnica Bucuresti, Romania;
Ministry of Education, Science, Research and Sport of the Slovak Republic, Slovakia;
National Research Foundation of South Africa, South Africa;
Swedish Research Council (VR) and Knut \& Alice Wallenberg Foundation (KAW), Sweden;
European Organization for Nuclear Research, Switzerland;
Suranaree University of Technology (SUT), National Science and Technology Development Agency (NSTDA) and National Science, Research and Innovation Fund (NSRF via PMU-B B05F650021), Thailand;
Turkish Energy, Nuclear and Mineral Research Agency (TENMAK), Turkey;
National Academy of  Sciences of Ukraine, Ukraine;
Science and Technology Facilities Council (STFC), United Kingdom;
National Science Foundation of the United States of America (NSF) and United States Department of Energy, Office of Nuclear Physics (DOE NP), United States of America.
In addition, individual groups or members have received support from:
European Research Council, Strong 2020 - Horizon 2020 (grant nos. 950692, 824093), European Union;
Academy of Finland (Center of Excellence in Quark Matter) (grant nos. 346327, 346328), Finland;
Programa de Apoyos para la Superaci\'{o}n del Personal Acad\'{e}mico, UNAM, Mexico.
\end{acknowledgement}

\bibliographystyle{utphys}   
\bibliography{bibliography}

\newpage
\appendix

%
%
\section{The ALICE Collaboration}
\label{app:collab}
\begin{flushleft} 
\small

S.~Acharya\,\orcidlink{0000-0002-9213-5329}\,$^{\rm 127}$, 
D.~Adamov\'{a}\,\orcidlink{0000-0002-0504-7428}\,$^{\rm 87}$, 
A.~Adler$^{\rm 71}$, 
G.~Aglieri Rinella\,\orcidlink{0000-0002-9611-3696}\,$^{\rm 33}$, 
M.~Agnello\,\orcidlink{0000-0002-0760-5075}\,$^{\rm 30}$, 
N.~Agrawal\,\orcidlink{0000-0003-0348-9836}\,$^{\rm 52}$, 
Z.~Ahammed\,\orcidlink{0000-0001-5241-7412}\,$^{\rm 135}$, 
S.~Ahmad\,\orcidlink{0000-0003-0497-5705}\,$^{\rm 16}$, 
S.U.~Ahn\,\orcidlink{0000-0001-8847-489X}\,$^{\rm 72}$, 
I.~Ahuja\,\orcidlink{0000-0002-4417-1392}\,$^{\rm 38}$
% ...

\section*{Affiliation Notes}

$^{\rm I}$ Deceased\\
$^{\rm II}$ Also at: Max-Planck-Institut fur Physik, Munich, Germany\\
$^{\rm III}$ Also at: Italian National Agency for New Technologies, Energy and Sustainable Economic Development (ENEA), Bologna, Italy\\
$^{\rm IV}$ Also at: Dipartimento DET del Politecnico di Torino, Turin, Italy\\
$^{\rm V}$ Also at: Department of Applied Physics, Aligarh Muslim University, Aligarh, India\\
$^{\rm VI}$ Also at: Institute of Theoretical Physics, University of Wroclaw, Poland\\
$^{\rm VII}$ Also at: An institution covered by a cooperation agreement with CERN\\

\section*{Collaboration Institutes}

$^{1}$ A.I. Alikhanyan National Science Laboratory (Yerevan Physics Institute) Foundation, Yerevan, Armenia\\
$^{2}$ AGH University of Krakow, Cracow, Poland\\
$^{3}$ Bogolyubov Institute for Theoretical Physics, National Academy of Sciences of Ukraine, Kiev, Ukraine\\
$^{4}$ Bose Institute, Department of Physics  and Centre for Astroparticle Physics and Space Science (CAPSS), Kolkata, India\\
$^{5}$ California Polytechnic State University, San Luis Obispo, California, United States\\
$^{6}$ Central China Normal University, Wuhan, China\\
$^{7}$ Centro de Aplicaciones Tecnol\'{o}gicas y Desarrollo Nuclear (CEADEN), Havana, Cuba\\
$^{8}$ Centro de Investigaci\'{o}n y de Estudios Avanzados (CINVESTAV), Mexico City and M\'{e}rida, Mexico\\
$^{9}$ Chicago State University, Chicago, Illinois, United States\\
$^{10}$ China Institute of Atomic Energy, Beijing, China\\
$^{11}$ China University of Geosciences, Wuhan, China\\
$^{12}$ Chungbuk National University, Cheongju, Republic of Korea\\
$^{13}$ Comenius University Bratislava, Faculty of Mathematics, Physics and Informatics, Bratislava, Slovak Republic\\
$^{14}$ COMSATS University Islamabad, Islamabad, Pakistan\\
$^{15}$ Creighton University, Omaha, Nebraska, United States\\
$^{16}$ Department of Physics, Aligarh Muslim University, Aligarh, India\\
$^{17}$ Department of Physics, Pusan National University, Pusan, Republic of Korea\\
$^{18}$ Department of Physics, Sejong University, Seoul, Republic of Korea\\
$^{19}$ Department of Physics, University of California, Berkeley, California, United States\\
$^{20}$ Department of Physics, University of Oslo, Oslo, Norway\\
$^{21}$ Department of Physics and Technology, University of Bergen, Bergen, Norway\\
$^{22}$ Dipartimento di Fisica, Universit\`{a} di Pavia, Pavia, Italy\\
$^{23}$ Dipartimento di Fisica dell'Universit\`{a} and Sezione INFN, Cagliari, Italy\\
$^{24}$ Dipartimento di Fisica dell'Universit\`{a} and Sezione INFN, Trieste, Italy\\
$^{25}$ Dipartimento di Fisica dell'Universit\`{a} and Sezione INFN, Turin, Italy\\
$^{26}$ Dipartimento di Fisica e Astronomia dell'Universit\`{a} and Sezione INFN, Bologna, Italy\\
$^{27}$ Dipartimento di Fisica e Astronomia dell'Universit\`{a} and Sezione INFN, Catania, Italy\\
$^{28}$ Dipartimento di Fisica e Astronomia dell'Universit\`{a} and Sezione INFN, Padova, Italy\\
$^{29}$ Dipartimento di Fisica `E.R.~Caianiello' dell'Universit\`{a} and Gruppo Collegato INFN, Salerno, Italy\\
$^{30}$ Dipartimento DISAT del Politecnico and Sezione INFN, Turin, Italy\\
$^{31}$ Dipartimento di Scienze MIFT, Universit\`{a} di Messina, Messina, Italy\\
$^{32}$ Dipartimento Interateneo di Fisica `M.~Merlin' and Sezione INFN, Bari, Italy\\
$^{33}$ European Organization for Nuclear Research (CERN), Geneva, Switzerland\\
$^{34}$ Faculty of Electrical Engineering, Mechanical Engineering and Naval Architecture, University of Split, Split, Croatia\\
$^{35}$ Faculty of Engineering and Science, Western Norway University of Applied Sciences, Bergen, Norway\\
$^{36}$ Faculty of Nuclear Sciences and Physical Engineering, Czech Technical University in Prague, Prague, Czech Republic\\
$^{37}$ Faculty of Physics, Sofia University, Sofia, Bulgaria\\
$^{38}$ Faculty of Science, P.J.~\v{S}af\'{a}rik University, Ko\v{s}ice, Slovak Republic\\
$^{39}$ Frankfurt Institute for Advanced Studies, Johann Wolfgang Goethe-Universit\"{a}t Frankfurt, Frankfurt, Germany\\
$^{40}$ Fudan University, Shanghai, China\\
$^{41}$ Gangneung-Wonju National University, Gangneung, Republic of Korea\\
$^{42}$ Gauhati University, Department of Physics, Guwahati, India\\
$^{43}$ Helmholtz-Institut f\"{u}r Strahlen- und Kernphysik, Rheinische Friedrich-Wilhelms-Universit\"{a}t Bonn, Bonn, Germany\\
$^{44}$ Helsinki Institute of Physics (HIP), Helsinki, Finland\\
$^{45}$ High Energy Physics Group,  Universidad Aut\'{o}noma de Puebla, Puebla, Mexico\\
$^{46}$ Horia Hulubei National Institute of Physics and Nuclear Engineering, Bucharest, Romania\\
$^{47}$ HUN-REN Wigner Research Centre for Physics, Budapest, Hungary\\
$^{48}$ Indian Institute of Technology Bombay (IIT), Mumbai, India\\
$^{49}$ Indian Institute of Technology Indore, Indore, India\\
$^{50}$ INFN, Laboratori Nazionali di Frascati, Frascati, Italy\\
$^{51}$ INFN, Sezione di Bari, Bari, Italy\\
$^{52}$ INFN, Sezione di Bologna, Bologna, Italy\\
$^{53}$ INFN, Sezione di Cagliari, Cagliari, Italy\\
$^{54}$ INFN, Sezione di Catania, Catania, Italy\\
$^{55}$ INFN, Sezione di Padova, Padova, Italy\\
$^{56}$ INFN, Sezione di Pavia, Pavia, Italy\\
$^{57}$ INFN, Sezione di Torino, Turin, Italy\\
$^{58}$ INFN, Sezione di Trieste, Trieste, Italy\\
$^{59}$ Inha University, Incheon, Republic of Korea\\
$^{60}$ Institute for Gravitational and Subatomic Physics (GRASP), Utrecht University/Nikhef, Utrecht, Netherlands\\
$^{61}$ Institute of Experimental Physics, Slovak Academy of Sciences, Ko\v{s}ice, Slovak Republic\\
$^{62}$ Institute of Physics, Homi Bhabha National Institute, Bhubaneswar, India\\
$^{63}$ Institute of Physics of the Czech Academy of Sciences, Prague, Czech Republic\\
$^{64}$ Institute of Space Science (ISS), Bucharest, Romania\\
$^{65}$ Institut f\"{u}r Kernphysik, Johann Wolfgang Goethe-Universit\"{a}t Frankfurt, Frankfurt, Germany\\
$^{66}$ Instituto de Ciencias Nucleares, Universidad Nacional Aut\'{o}noma de M\'{e}xico, Mexico City, Mexico\\
$^{67}$ Instituto de F\'{i}sica, Universidade Federal do Rio Grande do Sul (UFRGS), Porto Alegre, Brazil\\
$^{68}$ Instituto de F\'{\i}sica, Universidad Nacional Aut\'{o}noma de M\'{e}xico, Mexico City, Mexico\\
$^{69}$ iThemba LABS, National Research Foundation, Somerset West, South Africa\\
$^{70}$ Jeonbuk National University, Jeonju, Republic of Korea\\
$^{71}$ Johann-Wolfgang-Goethe Universit\"{a}t Frankfurt Institut f\"{u}r Informatik, Fachbereich Informatik und Mathematik, Frankfurt, Germany\\
$^{72}$ Korea Institute of Science and Technology Information, Daejeon, Republic of Korea\\
$^{73}$ KTO Karatay University, Konya, Turkey\\
$^{74}$ Laboratoire de Physique Subatomique et de Cosmologie, Universit\'{e} Grenoble-Alpes, CNRS-IN2P3, Grenoble, France\\
$^{75}$ Lawrence Berkeley National Laboratory, Berkeley, California, United States\\
$^{76}$ Lund University Department of Physics, Division of Particle Physics, Lund, Sweden\\
$^{77}$ Nagasaki Institute of Applied Science, Nagasaki, Japan\\
$^{78}$ Nara Women{'}s University (NWU), Nara, Japan\\
$^{79}$ National and Kapodistrian University of Athens, School of Science, Department of Physics , Athens, Greece\\
$^{80}$ National Centre for Nuclear Research, Warsaw, Poland\\
$^{81}$ National Institute of Science Education and Research, Homi Bhabha National Institute, Jatni, India\\
$^{82}$ National Nuclear Research Center, Baku, Azerbaijan\\
$^{83}$ National Research and Innovation Agency - BRIN, Jakarta, Indonesia\\
$^{84}$ Niels Bohr Institute, University of Copenhagen, Copenhagen, Denmark\\
$^{85}$ Nikhef, National institute for subatomic physics, Amsterdam, Netherlands\\
$^{86}$ Nuclear Physics Group, STFC Daresbury Laboratory, Daresbury, United Kingdom\\
$^{87}$ Nuclear Physics Institute of the Czech Academy of Sciences, Husinec-\v{R}e\v{z}, Czech Republic\\
$^{88}$ Oak Ridge National Laboratory, Oak Ridge, Tennessee, United States\\
$^{89}$ Ohio State University, Columbus, Ohio, United States\\
$^{90}$ Physics department, Faculty of science, University of Zagreb, Zagreb, Croatia\\
$^{91}$ Physics Department, Panjab University, Chandigarh, India\\
$^{92}$ Physics Department, University of Jammu, Jammu, India\\
$^{93}$ Physics Program and International Institute for Sustainability with Knotted Chiral Meta Matter (SKCM2), Hiroshima University, Hiroshima, Japan\\
$^{94}$ Physikalisches Institut, Eberhard-Karls-Universit\"{a}t T\"{u}bingen, T\"{u}bingen, Germany\\
$^{95}$ Physikalisches Institut, Ruprecht-Karls-Universit\"{a}t Heidelberg, Heidelberg, Germany\\
$^{96}$ Physik Department, Technische Universit\"{a}t M\"{u}nchen, Munich, Germany\\
$^{97}$ Politecnico di Bari and Sezione INFN, Bari, Italy\\
$^{98}$ Research Division and ExtreMe Matter Institute EMMI, GSI Helmholtzzentrum f\"ur Schwerionenforschung GmbH, Darmstadt, Germany\\
$^{99}$ Saga University, Saga, Japan\\
$^{100}$ Saha Institute of Nuclear Physics, Homi Bhabha National Institute, Kolkata, India\\
$^{101}$ School of Physics and Astronomy, University of Birmingham, Birmingham, United Kingdom\\
$^{102}$ Secci\'{o}n F\'{\i}sica, Departamento de Ciencias, Pontificia Universidad Cat\'{o}lica del Per\'{u}, Lima, Peru\\
$^{103}$ Stefan Meyer Institut f\"{u}r Subatomare Physik (SMI), Vienna, Austria\\
$^{104}$ SUBATECH, IMT Atlantique, Nantes Universit\'{e}, CNRS-IN2P3, Nantes, France\\
$^{105}$ Sungkyunkwan University, Suwon City, Republic of Korea\\
$^{106}$ Suranaree University of Technology, Nakhon Ratchasima, Thailand\\
$^{107}$ Technical University of Ko\v{s}ice, Ko\v{s}ice, Slovak Republic\\
$^{108}$ The Henryk Niewodniczanski Institute of Nuclear Physics, Polish Academy of Sciences, Cracow, Poland\\
$^{109}$ The University of Texas at Austin, Austin, Texas, United States\\
$^{110}$ Universidad Aut\'{o}noma de Sinaloa, Culiac\'{a}n, Mexico\\
$^{111}$ Universidade de S\~{a}o Paulo (USP), S\~{a}o Paulo, Brazil\\
$^{112}$ Universidade Estadual de Campinas (UNICAMP), Campinas, Brazil\\
$^{113}$ Universidade Federal do ABC, Santo Andre, Brazil\\
$^{114}$ Universitatea Nationala de Stiinta si Tehnologie Politehnica Bucuresti, Bucharest, Romania\\
$^{115}$ University of Cape Town, Cape Town, South Africa\\
$^{116}$ University of Houston, Houston, Texas, United States\\
$^{117}$ University of Jyv\"{a}skyl\"{a}, Jyv\"{a}skyl\"{a}, Finland\\
$^{118}$ University of Kansas, Lawrence, Kansas, United States\\
$^{119}$ University of Liverpool, Liverpool, United Kingdom\\
$^{120}$ University of Science and Technology of China, Hefei, China\\
$^{121}$ University of South-Eastern Norway, Kongsberg, Norway\\
$^{122}$ University of Tennessee, Knoxville, Tennessee, United States\\
$^{123}$ University of the Witwatersrand, Johannesburg, South Africa\\
$^{124}$ University of Tokyo, Tokyo, Japan\\
$^{125}$ University of Tsukuba, Tsukuba, Japan\\
$^{126}$ Universit\"{a}t M\"{u}nster, Institut f\"{u}r Kernphysik, M\"{u}nster, Germany\\
$^{127}$ Universit\'{e} Clermont Auvergne, CNRS/IN2P3, LPC, Clermont-Ferrand, France\\
$^{128}$ Universit\'{e} de Lyon, CNRS/IN2P3, Institut de Physique des 2 Infinis de Lyon, Lyon, France\\
$^{129}$ Universit\'{e} de Strasbourg, CNRS, IPHC UMR 7178, F-67000 Strasbourg, France, Strasbourg, France\\
$^{130}$ Universit\'{e} Paris-Saclay, Centre d'Etudes de Saclay (CEA), IRFU, D\'{e}partment de Physique Nucl\'{e}aire (DPhN), Saclay, France\\
$^{131}$ Universit\'{e}  Paris-Saclay, CNRS/IN2P3, IJCLab, Orsay, France\\
$^{132}$ Universit\`{a} degli Studi di Foggia, Foggia, Italy\\
$^{133}$ Universit\`{a} del Piemonte Orientale, Vercelli, Italy\\
$^{134}$ Universit\`{a} di Brescia, Brescia, Italy\\
$^{135}$ Variable Energy Cyclotron Centre, Homi Bhabha National Institute, Kolkata, India\\
$^{136}$ Warsaw University of Technology, Warsaw, Poland\\
$^{137}$ Wayne State University, Detroit, Michigan, United States\\
$^{138}$ Yale University, New Haven, Connecticut, United States\\
$^{139}$ Yonsei University, Seoul, Republic of Korea\\
$^{140}$  Zentrum  f\"{u}r Technologie und Transfer (ZTT), Worms, Germany\\
$^{141}$ Affiliated with an institute covered by a cooperation agreement with CERN\\
$^{142}$ Affiliated with an international laboratory covered by a cooperation agreement with CERN.\\

\end{flushleft} 

\end{document}